\documentclass[a4paper,nofootinbib,preprintnumbers,twocolumn,preprintnumbers,floatfix,superscriptaddress,pra,showpacs]{revtex4-1}

\usepackage{graphicx}
\usepackage{amssymb}
\usepackage{amsmath}
\usepackage{epstopdf}
\usepackage{graphics}
\usepackage[caption=false]{subfig}
\usepackage{float}
\usepackage{color}
\usepackage{hyperref}
\usepackage{bbold}

\allowdisplaybreaks[1]

\newcommand{\bra}[1]{\langle #1|}					
\newcommand{\ket}[1]{|#1\rangle}					

\newcommand{\figref}[1]{Fig.~\ref{#1}}
\newcommand{\tabref}[1]{Table \ref{#1}}
\newcommand{\secref}[1]{Sec.~\ref{#1}}

\begin{document}

\title{Long-distance entanglement distribution using individual atoms in optical cavities}

\author{J. Borregaard }
\affiliation{The Niels Bohr Institute, University of Copenhagen, Blegdamsvej 17, DK-2100 Copenhagen \O, Denmark}
\affiliation{Department of Physics, Harvard University, Cambridge, MA 02138, USA}
\author{P. K\'om\'ar}
\affiliation{Department of Physics, Harvard University, Cambridge, MA 02138, USA}
\author{E. M. Kessler} 
\affiliation{Department of Physics, Harvard University, Cambridge, MA 02138, USA}
\affiliation{ITAMP, Harvard-Smithsonian Center for Astrophysics, Cambridge, MA 02138, USA}
\author{M. D. Lukin}
\affiliation{Department of Physics, Harvard University, Cambridge, MA 02138, USA}
\author{A. S. S\o rensen}
\affiliation{The Niels Bohr Institute, University of Copenhagen, Blegdamsvej 17, DK-2100 Copenhagen \O, Denmark}

\date{\today}

\begin{abstract}
Individual atoms in optical cavities can provide an efficient interface between stationary qubits and flying qubits (photons), which is an essentiel building block for quantum communication. Furthermore, cavity assisted controlled-not (CNOT) gates can be used for swapping entanglement to long distances in a quantum repeater setup. Nonetheless, dissipation introduced by the cavity during the CNOT may increase the experimental difficulty in obtaining long-distance entanglement distribution using these systems. We analyse and compare a number of cavity-based repeater schemes combining various entanglement generation schemes and cavity assisted CNOT gates. We find that a scheme, where high-fidelity entanglement is first generated in a two-photon detection scheme and then swapped to long distances using a recently proposed heralded CZ-gate exhibits superior performance compared to the other schemes. The heralded gate moves the effect of dissipation from the fidelity to the success probability of the gate thereby enabling high-fidelity entanglement swapping. As a result, high-rate entanglement distribution can be achieved over long distances even for low cooperativities of the atom-cavity systems. This high-fidelity repeater is shown to outperform the other cavity-based schemes by up to two orders of magnitude in the rate for realistic parameters and large distances ($1000$ km).   
\end{abstract}


\maketitle

\section{Introduction}

Distribution of entanglement is an essential task in quantum communication~\cite{kimble,cirac, acin}.  Entanglement can be used to make highly secure communication channels due to the sensitivity of entangled quantum systems to external influences~\cite{scarani}. While this sensitivity makes it possible to detect any attack from an eavesdropper, it also makes it hard to distribute entanglement over large distances since any noise from the enviroment quickly destroyes the entanglement. Direct transmission of a quantum signal suffers from loss and decoherence from the transmission channel, which results in an exponential decrease of the rate with distance~\cite{briegel}. To overcome this problem, it has been proposed to use quantum repeaters, where entanglement is first created over short distances by direct transmission and then stored in quantum memories until it can be swapped to larger distances~\cite{briegel,duan3} (See \figref{fig:figure1}). Much effort has been devoted to the construction of quantum repeaters based on atomic ensembles, where the large number of atoms, in principle, enables highly efficient quantum memories~\cite{sangouard3,cell}. Nonetheless, the limited efficiencies demonstrated in current experiments with atomic ensembles~\cite{sangouard3,hammerer} prevents the construction of a practical quantum repeater based on currently existing setups.

Single emitter systems such as color centers and trapped ions have also been considered for quantum repeaters~\cite{childress,sangouard2}. The long coherence times demonstrated with, e.g. trapped ions make them desirable as quantum memories. Nonetheless, entanglement needs to be created non-locally between two memories in the initial step of a repeater. This requires efficient transfer of information from the quantum memories onto light in the form of single photons. To this end, it is an advantage to place the emitter inside a cavity, which can greatly enhance the light-emitter coupling~\cite{acin,ritter}. Entanglement swapping can then be performed with a cavity mediated CNOT gate ~\cite{zoller1,haroche1} but in this case, the detrimental effect of cavity loss and spontaneous emission from the emitters may prevent obtaining efficient entanglement swapping. The parameter characterizing the effect of dissipation in the emitter-cavity system is the cooperativity, $C$. It has been argued that a direct implementation of gates in a cavity will make the gate fidelity, $F$, have a poor scaling of $F\sim1-1/\sqrt{C}$ \cite{kastoryano,Anders2prl}. To overcome this problem for current cavities with limited $C$, it has been suggested to employ entanglement purification after each swap operation to boost the entanglement but this either requires a large number of resources or a time consuming sequential generation of purification pairs~\cite{bennett,deutsch,duan4,pan}. 

Here we analyze and compare a number of cavity-based quantum repeaters which combines various proposals for entanglement generation and cavity-assisted CNOT gates. We find that the best scheme is where high-fidelity entanglement is generated using a two-photon detection scheme similar to Ref.~\cite{kimble2} and swapped to large distances using the heralded CZ-gate proposed in Ref.~\cite{johannes}. The heralded gate enables nearly perfect entanglement swapping when successful allowing for many swaps without the need of entanglement purification. As a result, high-rate entanglement distribution is achieved even for low cooperativities. 

Compared to the other cavity-based repeaters, this high-fidelity repeater achieves up to two orders of magnitude higher secret key rate (see below) for realistic parameters and large distances ($1000$ km). Specifically, we have compared to repeaters where entanglement is generated using a single-photon detection scheme similar to Ref.~\cite{huelga}, which allows for a better rate at the expense of fidelity. Furthermore we have considered schemes where entanglement swapping is achieved using the deterministic CNOT gate suggested in Ref.~\cite{Anders2prl}, combining it with the local entanglement generation scheme of Ref.~\cite{Anders1prl}. The advantage of this gate is that the fidelity scales as $F\sim1-1/C$, which is a significant improvement of the $1/\sqrt{C}$ scaling characterizing the performance of a direct implementation of gates in a cavity. As a result, long-distance entanglement distribution can also be achieved with this gate but it requires cooperativities above $100$, which might be challenging to achieve experimentally. Furthermore, we include the possibility of initial purification in repeaters based on the single-photon detection scheme in order to allow for the higher rate of this scheme to compensate for the lower fidelity compared to the two-photon detection scheme. 

To reflect a realistic near-term approach to quantum repeaters, we only consider scenarios with 2 or 4 qubits per repeater station. For the same reason, we also do not consider the possibility of intermediate entanglement purification. Here, initial purification refers to purification in the elementary links (see \figref{fig:figure1}) while intermediate purification refers to purification in the subsequent stages of a repeater. We have numerically optimized all the considered repeater schemes for a range of cooperativities and distances to find the highest achievable secret key rate (see below). Note that a similar optimization of repeater schemes based on dynamical programming was described in Ref.~\cite{jiang2007}. In that work, both initial and intermediate entanglement purification was considered assuming high-fidelity operations.  Our optimization is less detailed since we do not consider intermediate purification. On the other hand, we include how the errors of the operations depend on the physical parameters such as the cooperativity and investigate concrete physical implementations. 
Finally, we compare the high-fidelity repeater considered here to both an ion-trap repeater and one of the best repeaters based on atomic ensembles~\cite{sangouard3}. For a distance of 1000 km, the high-fidelity repeater outperforms both of these schemes for $C\gtrsim30$.       

\section{High-fidelity quantum repeater}  \label{sec:generation}

\begin{figure} 
\centering
\includegraphics[width=3.0in, height=1.5in]{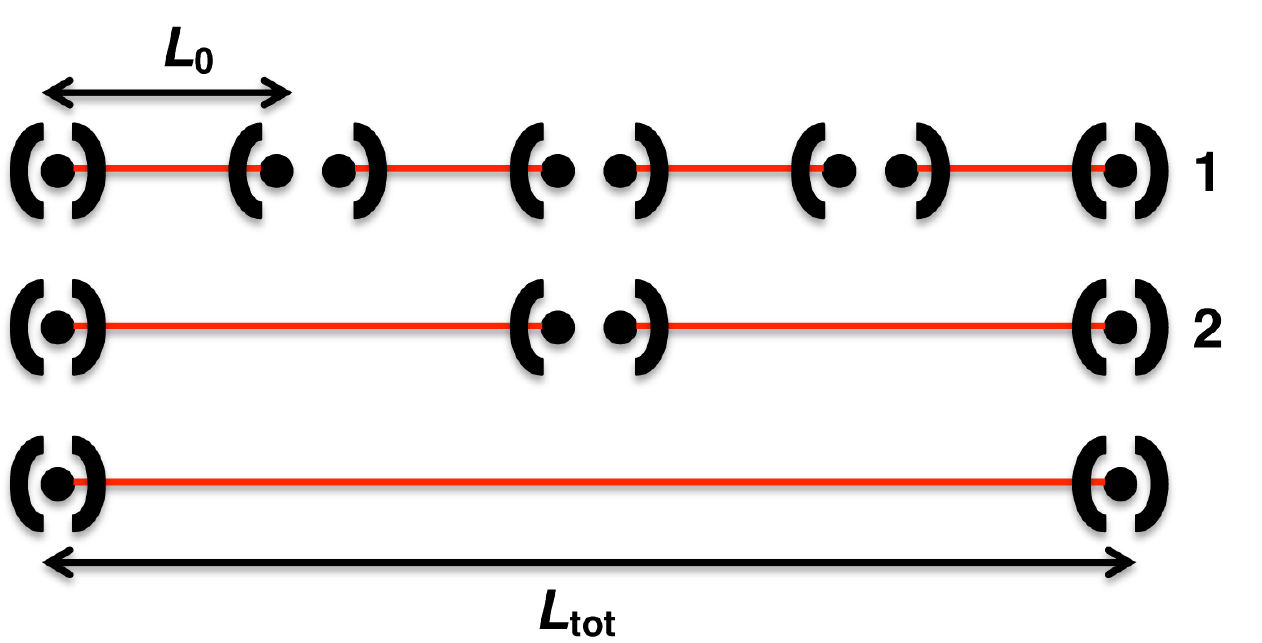}
\caption{The general architecture of a quantum repeater. The total distance, over which entanglement should be distributed, is divided into elementary links of length $L_{0}$ connected by repeater stations pictured as cavities containing single emitters. After creating entanglement in the elementary links the entanglement is swapped to larger distances by combining the elementary links. The numbers to the right in the figure refers to the swap level of the repeater. In the first swap level, the four elementary links are connected pairwise to make two longer links. In the second swap levels these two links are connected to create entanglement over the total distance. The total number of swap levels is thus 2 for this depicted setup. }
\label{fig:figure1}
\end{figure}

\begin{figure} 
\centering
\includegraphics[width=0.5\textwidth]{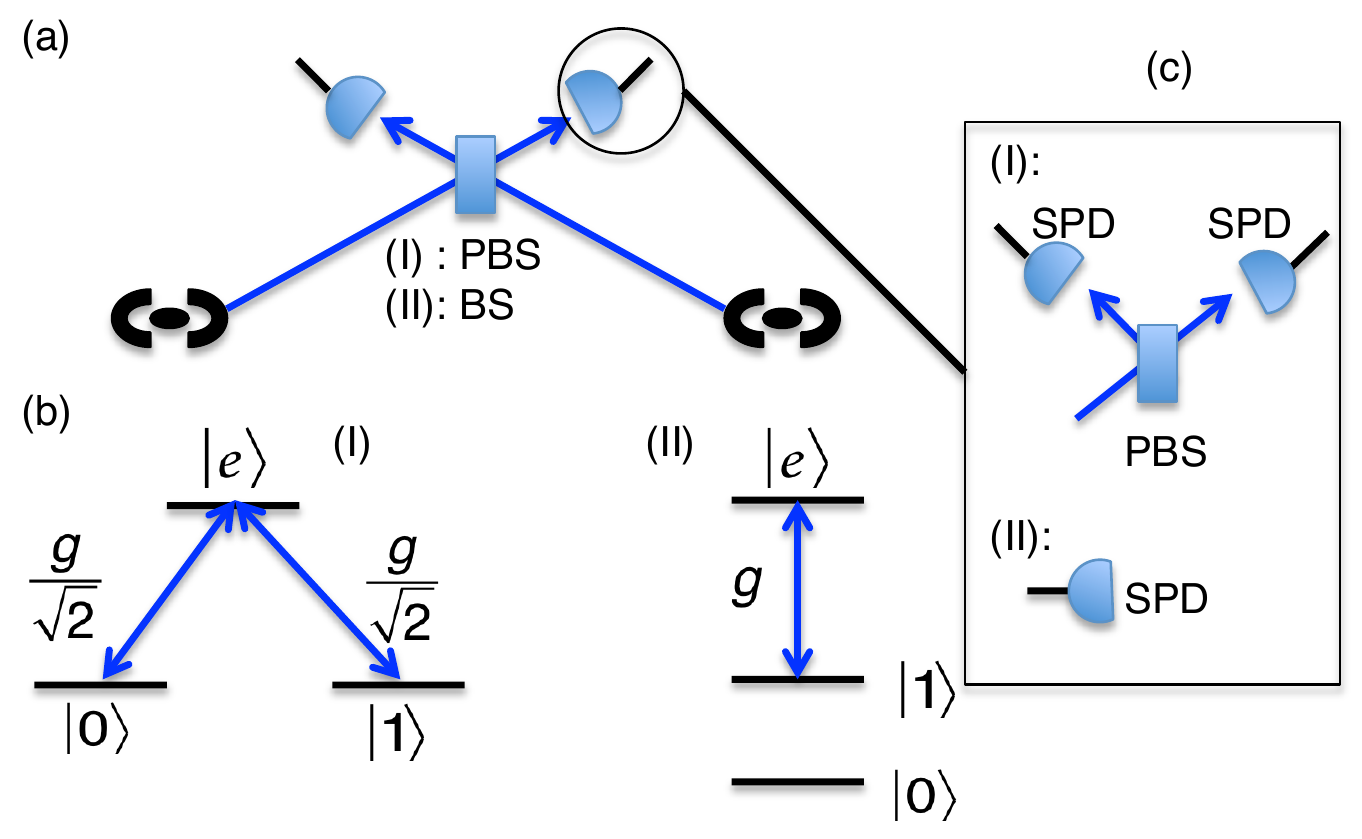}
\caption[Entanglement generation I]{Entanglement generation in the elementary links where emission from two cavities are combined on a beam splitter. (a) shows the basic setup, (b) shows the level structure of the emitters and (c) shows the detection setup.  (I) refers to the two-photon detection scheme and (II) refers to the single-photon detection scheme. Both schemes use a central station with either (I) three polarizing beam splitters (PBS) and four single-photon detectors or (II) a single balanced beam splitter (BS) and two single-photon detectors.  $g$ denotes the cavity coupling. For the two-photon scheme the levels $\ket{0}$ and $\ket{1}$ are assumed to have equal coupling of $g/\sqrt{2}$ to the excited state $\ket{e}$.}
\label{fig:figure2}
\end{figure}

We will first describe the details of the high-fidelity quantum repeater, which we find to have the best performance and later discuss and compare with the various other schemes. The first step in any quantum repeater is to create non-local entanglement in the elementary links (see \figref{fig:figure1}).
To this end, a two-photon detection scheme, as proposed in Ref.~\cite{kimble2}, is considered. The basic setup is shown in \figref{fig:figure2}(a).
Both emitters are initially prepared in the excited state $\ket{e}$ by a strong excitation pulse and the cavity is assumed to couple both the $\ket{e}\to\ket{1}$ and $\ket{e}\to\ket{0}$ transitions with equal coupling strength $g/\sqrt{2}$ (see \figref{fig:figure2}(b)). The two transitions are, however, assumed to produce photons with different polarizations such that the emission of a cavity photon creates an entangled state between the photon and the emitter of the form $\frac{1}{\sqrt{2}}\left(\ket{0}\ket{1_{1}}_{L}+\ket{1}\ket{1_{2}}_{L}\right)$ where $\ket{1_{1}}_{L}$ ($\ket{1_{2}}_{L}$) is the single photon state with polarization 1 (2). The probability of one of the emitters to emit a photon of either polarization through the cavity, into an optical fiber transmitting it to the detection stage during a time interval $[0;T]$ is  
\begin{equation} \label{eq:phot1}
P_{\text{phot}}=\frac{4C}{1+4C}\left(1-e^{-\gamma(1+4C)T}\right), 
\end{equation} 
assuming perfect outcoupling to the fiber and that the decay rate of the cavity, $\kappa$, is much larger than the cavity coupling, $g$. We have here introduced the cooperativity $C=g^{2}/\kappa\gamma$, where $\gamma$ is the spontaneous emission rate of the emitters into modes other than the cavity. This is the key parameter characterizing the performance of the cavity-based repeaters. The photons are sent from the cavities to a central polarizing beam splitter (PBS). If two photons of the same polarizations are incident on the PBS, they leave in different output ports, while photons of different polarization leave in the same output port. The outputs are then sent to a second set of polarizing beam splitters and all four outputs of these are finally measured with single photon detectors (SPD). A click in a detector in each arm heralds the creation of the Bell state $\ket{\Phi^{+}}=(\ket{00}+\ket{11})/\sqrt{2}$ between the emitters up to a local qubit rotation. Neglecting dark counts of the detectors, the heralded fidelity is unity (see App.~\ref{two}) while the success probability of the scheme is $P_{\text{2click}}=\frac{1}{2}\eta^{2}P_{\text{phot}}^{2}$ with $\eta$ being the total detection efficiency including inefficient outcoupling of the cavity light, losses in the transmission fibers and imperfect detectors. Compared to schemes based on single-photon detection (see below) the rate of this two-photon detection scheme decreases rapidly with decreasing $\eta$. On the other hand it has a high fidelity, which is desirable for the subsequent stages of entanglement swapping as we will show below. 

\begin{figure}
\centering
\subfloat {\label{fig:figure3a}\includegraphics[width=0.17\textwidth]{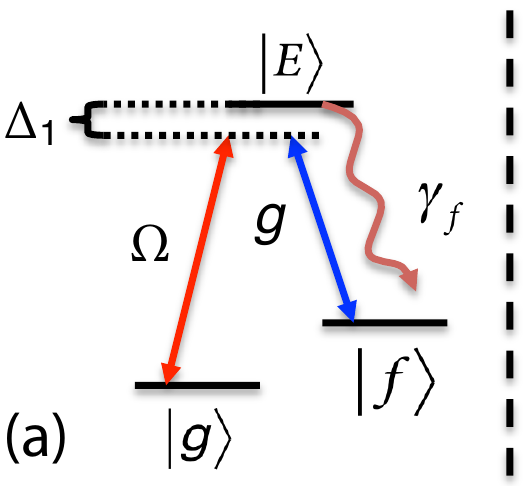}} 
\subfloat{\label{fig:figure3b}\includegraphics[width=0.15\textwidth]{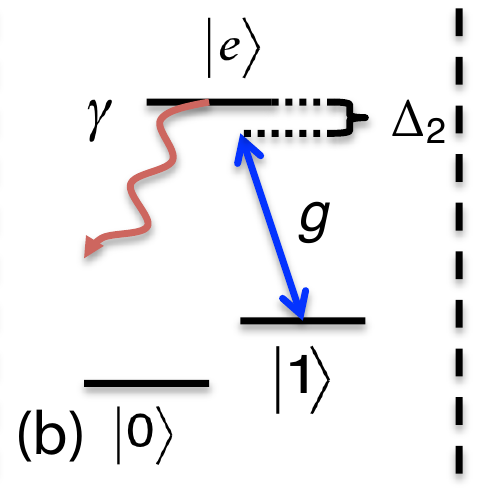}}
\subfloat{\label{fig:figure3c}\includegraphics[width=0.17\textwidth]{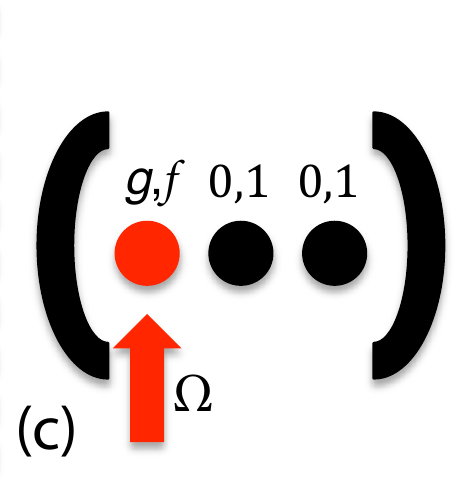}}
\caption{ Schematics of the heralded CZ gate~\cite{johannes}. (a) is the level structure of the auxiliary atom, (b) is the level structure of the qubit atoms and (c) shows the cavity containing the auxiliary atom and the two qubit atoms. Assuming that $\ket{E}$ only decays to $\ket{f}$ by e.g. driving the transition $\ket{g}\to\ket{E}$ with a two photon process, any spontaneous emission or cavity decay will change the state of the auxiliary atom from the initial state $\ket{g}$ to $\ket{f}$. The gate is thus conditioned on measuring the auxliary atom in state $\ket{g}$ at the end of the gate.}
\label{fig:figure3}
\end{figure}

For entanglement swapping we find that the best performance is achieved using the heralded CZ-gate described in Ref.~\cite{johannes}. The gate was described in detail for ${}^{87}$Rb atoms in Ref.~\cite{johannes} but it can be easily generalized to any set of emitters, which have the appropriate level structure (see \figref{fig:figure3}). Note that the gate operation relies on only qubit state $\ket{1}$ coupling to the cavity while the states $\ket{0}$ and $\ket{1}$ had equal cavity couplings in the entanglement generation scheme. To achieve this change in couplings, the state $\ket{0}$ should be mapped to another level in between the entanglement generation and the gate operation. For a realization with alkali atoms where the qubit states would be realized in the hyperfine grounds states, this could be achieved by applying a magnetic field to resolve the hyperfine states and applying a microwave pulse resonant with only the $\ket{0}$ state. 

In the heralded gate, the cavity is assumed to contain two qubit atoms and one auxiliary atom to facilitate the gate. The auxiliary atom is initialized in a state $\ket{g}$ that does not couple to the cavity and it would therefore not interfere with the entanglement generation scheme. By addressing the auxiliary atom with a weak laser pulse, an AC Stark shift is introduced, which gives a phase that depends on the state of the qubit atoms. Together with single qubit rotations, this enables a CZ-gate between the two qubit atoms. Furthermore, the auxiliary atom can function as an error detector in the sense that any cavity decay or spontaneous emission changes the state of the atom. Performing a heralding measurement of the state of the auxiliary atom at the end of the driving pulse removes all dissipative errors. As a consequence, the gate gets limited only by non-adiabatic effects. As shown in Ref.~\cite{johannes}, a heralded error below $4\cdot10^{-5}$ is possible with a gate time of $\sim377/(\gamma\sqrt{C})$, where $\gamma$ is the atomic linewidth. The failure probability of the gate scales as $1/\sqrt{C}$ and the high fidelity thus comes at the cost of a finite but possibly low failure probability. A CZ gate combined with single qubit rotations is sufficient to perform direct entanglement swapping. For simplicity, we assume perfect single qubit rotations and 100\% efficient measurement of atomic states for all schemes considered. Relaxing this assumption will in general decrease the rate of all the considered repeater schemes but schemes with a high number of swap levels like the high-fidelity repeater will be influenced more on the rate than schemes with a low number of swap levels.     

The advantage of the high-fidelity repeater can be understood by considering the requirement for reaching a certain threshold fidelity, $F_{\text{final}}$ of the distributed pair. In this case, the maximum number of swap levels is $N_{max}\sim -\log_{2}(F_{\text{final}}/(\epsilon_{0}+\epsilon_{g}))$, where $\epsilon_{0},\epsilon_{g}\ll1$ are the errors of the initial entanglement generation and the entanglement swapping respectively. The combination of the high-fidelity two-photon detection scheme and the heralded gate thus makes it possible to have a repeater with many elementary links while maintaining a high fidelity of the final distributed pair even for low cooperativities since the error of the heralded gate is still high in this regime.

\subsection{Secret key rate} \label{sec:secret}
\begin{figure} 
\centering
\includegraphics[width=0.45\textwidth]{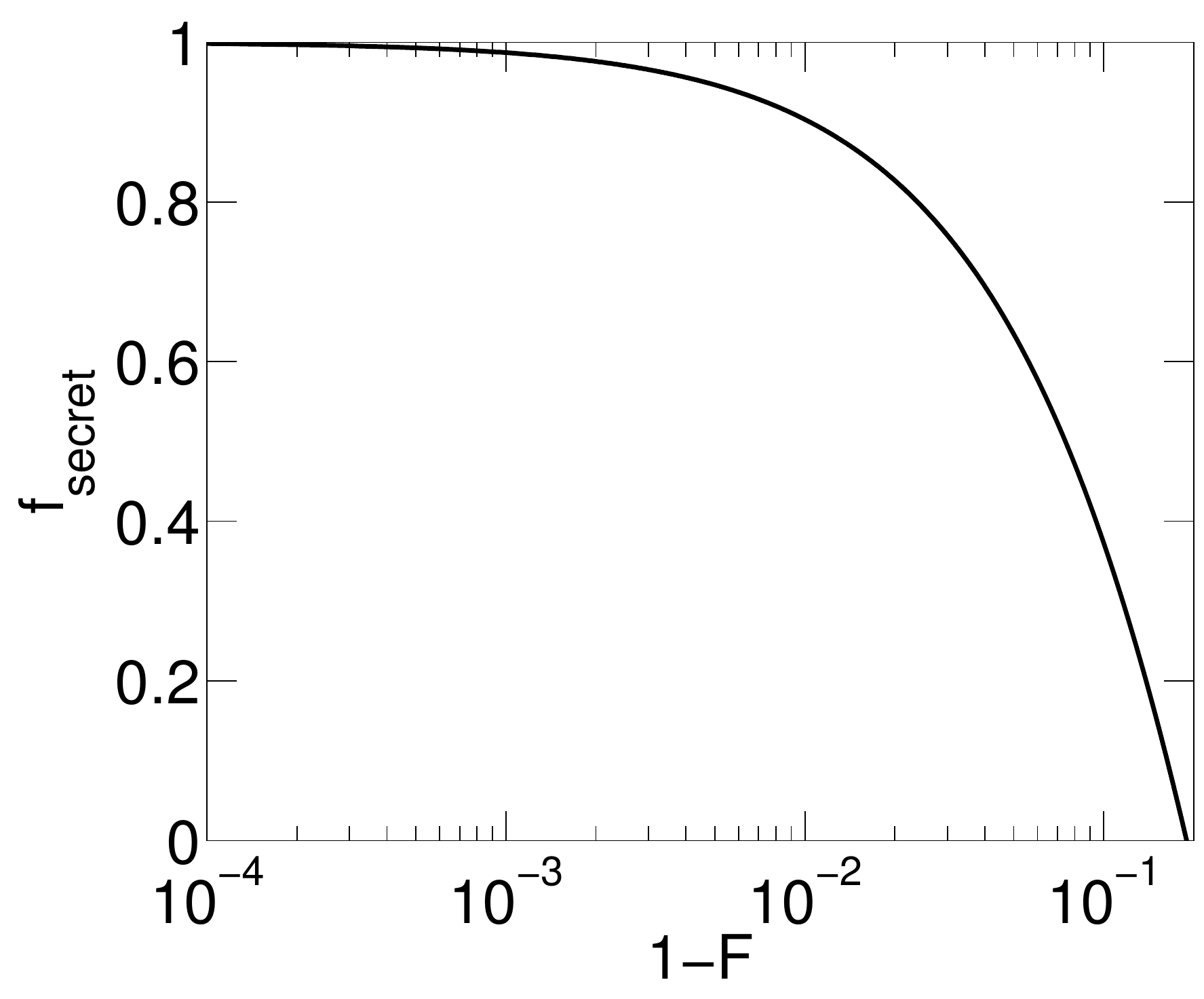}
\caption[Secret key fraction]{Secret key fraction ($f_{\text{secret}}$) as a function of the infidelity, $1-F$, of the final entangled pair. For $1-F\gtrsim19 \%$ it is no longer possible to extract a secret key from the raw keys.}
\label{fig:figure5}
\end{figure}

We imagine that the distributed entanglement is used to generate a secret key between two parties referred to as Alice and Bob. There exist various quantum key distribution schemes \cite{scarani,bennett2,ekert,bruss}, however, the general idea is that Alice and Bob can exclude that an eavesdropper has any information about the key by measuring their qubits and compairing results. We will assume that a six-state version of the BB84 protocol described in Ref. \cite{bruss} is used to generate the secret key. This protocol consists of three main steps. First Alice and Bob picks a basis according to some probability distribution and measure the state of their qubits thereby producing two bit strings referred to as the \emph{raw key}. Afterwards they compare their choice of basis and only keep the bits where they chose the same measurement basis thereby producing a \emph{sifted key}. Finally, Alice and Bob estimate the information that some eavesdropper could possibly have obtained about their key and perform privacy amplification \cite{scarani}. If the errors are not too big, they can obtain a shorter but completely secure key.  For the six-state protocol, the secret key rate, $r_{\text{secret}}$ can be defined as
\begin{equation}
r_{\text{secret}}=r_{\text{dist}}p_{\text{sift}}f_{\text{secret}},
\end{equation}
where $r_{\text{dist}}$ is the distribution rate of the entangled pairs, $p_{\text{sift}}$ is the probability that Alice and Bob choose the same measurement basis and $f_{\text{secret}}$ is the secret key fraction, which depends on the fidelity of the distributed pairs. We assume a worst case scenario where the distributed pairs are Werner states of the form 
\begin{eqnarray}
\rho&=&F\ket{\Phi^{+}}\bra{\Phi^{+}} \nonumber+ \frac{1-F}{3}\Big(\ket{\Phi^{-}}\bra{\Phi^{-}} \\&&+\ket{\Psi^{+}}\bra{\Psi^{+}}+\ket{\Psi^{-}}\bra{\Psi^{-}}\Big).
\end{eqnarray}
For such states, it is shown in Ref. \cite{scarani} that the secret key fraction in the six-state protocol can be estimated in the limit of infinitely long raw keys to be 
\begin{equation} \label{eq:secret2}
f_{\text{secret}}=1-h(\epsilon)-\epsilon+(1-\epsilon)h\left(\frac{1-3\epsilon/2}{1-\epsilon}\right),
\end{equation} 
where $\epsilon=2(1-F)/3$ and $h(p)=-p\text{log}_{2}(p)-(1-p)\text{log}_{2}(1-p)$ is the binary entropy. Eq.~\eqref{eq:secret2} is valid in the limit of perfect sifting and privacy amplification, which we assume to be the case. Furthermore, we assume an asymmetric version of the six state protocol, where one basis is used almost all the time such that $p_{\text{sift}}\approx1$ \cite{scarani}. \figref{fig:figure5} shows how the secret key fraction depends on the fidelity of the distributed pairs. 
As shown in the figure, high-fidelity pairs are required in order to have a non-vanishing secret key fraction. Again this points to the high-fidelity two-photon detection scheme and the nearly error-free heralded entanglement swapping as the best choice for the repeater.

\subsection{Repeater architecture}

\begin{figure} 
\centering
\includegraphics[width=0.48\textwidth]{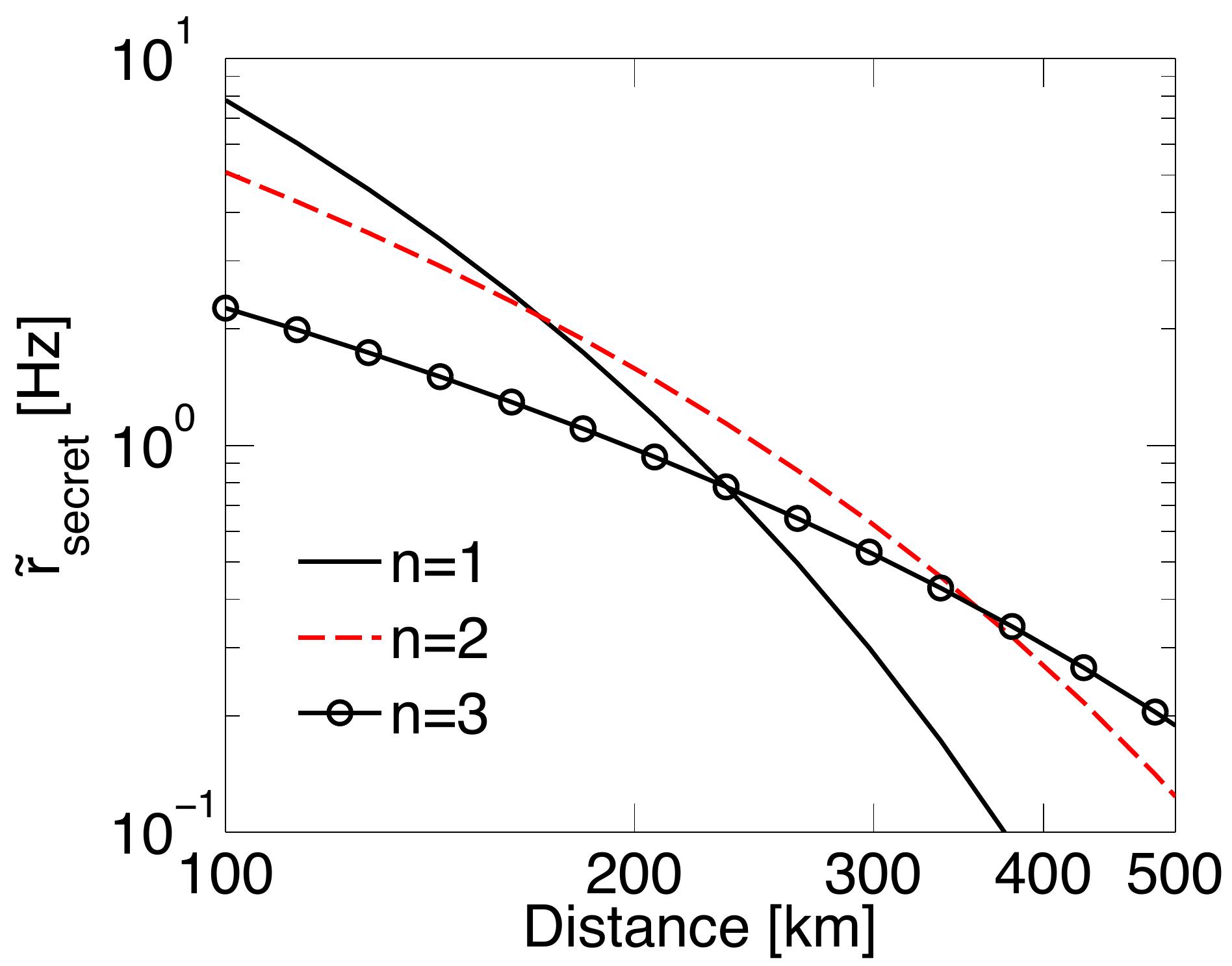}
\caption[Number of swap levels]{Normalized secret key rate per station($\tilde{r}_{\text{secret}}$) as a function of the distribution distance for a high-fidelity repeater consisting of the two-photon entanglement generation scheme and the heralded gate for entanglement swapping.  We have considered $n=2,3$ and 4 swap levels and have assumed a cooperativity of $C=$100 and two qubits per repeater station. The secret key rate was calculated as described in App.~\ref{app:rate} with the assumptions summarized in \tabref{tab:parameter}.  }
\label{fig:figureX4}
\end{figure}

The main goal of the quantum repeater is to overcome the effect of fiber losses. We model the fiber losses with a transmission efficiency $\eta_{\text{f}}=e^{-L_{0}/2L_{att}}$, where $L_{0}$ is the length of the elementary links of the repeater and $L_{att}$ is the fiber attenuation length. $\eta_{\text{f}}$ enters in the total detection efficiency $\eta$ as described above. For a given resource of $2^{n}+1$ repeater stations, one can either use all stations in a single repeater with $n$ swap levels or one can construct a number of parallel, independently operated chains of repeaters with less swap levels. Increasing the number of swap levels, decreases the fiber losses in the elementary links and thus increases the rate of entanglement generation. If, however, the length of the elementary links is already small, such that, e.g. imperfect SPD dominates the rate, then increasing the number of swap levels does not lead to any improvement. In this case it is advantageous to use the extra repeater stations to make another repeater with less swap levels, which runs in parallel with the already existing one. To make a proper assessment of the performance of repeater one should therefore include that adding swap levels costs resources in the form of additional repeater stations. In our comparison, we therefore consider a normalized secret key rate, $\tilde{r}_{secret}=r_{secret}$/(~\#~of stations), which is the secret key rate divided by the total number of repeater station instead of the bare secret key rate. To evaluate the performance of the repeater we calculate the achievable rate $\tilde r_{secret}$ as described in Appendix~\ref{app:rate} with the assumptions summarized in \tabref{tab:parameter} about fiber losses etc.. The resulting rate for various swap level used in the repeater is shown in \figref{fig:figureX4} as a function of distance. As seen in the figure, the optimal number of swap level changes with distance while considering the normalized secret key rate. The rate was calculated as described in App.~\ref{app:rate} for a cooperativity of 100 with the assumptions summarized in \tabref{tab:parameter} about fiber losses, detection efficiencies etc. For distances $\lesssim 150$ km, only a single swap level is needed since the fiber losses are relatively small while more swap levels are needed as the distance increases.   

In most repeater schemes, the qubits in each repeater station are assumed to be operated simultaneously with half of the qubits being used to generate entanglement with the neighbouring station to each side such that entanglement attempts in all the elementary links are done simultaneously. We will refer to this as a \emph{parallel} repeater.  We will, however, also consider another sequential way of operating the qubits, where all qubits in a station are first used to make entanglement in one elementary link. After this has been obtained, all but one qubit are then used to make entanglement over the neighboring link in the opposite direction. This is referred to as a \emph{sequential} repeater. The advantage of the sequential repeater is that the rate of the lowest level in the repeater, the entanglement generation, is increased. This comes at the cost of a waiting time between entanglement attempts in neighboring links. As the number of qubits per repeater station increases the sequential repeater will start to outperform the parallel repeater. We find that this happens with 4 qubits per repeater station (see Sec.~\ref{sec:optim}).     

\section{Other cavity based repeaters} \label{sec:other} 

We have found that the high-fidelity repeater that we have described above outperforms a number of other cavity-based repeater schemes, which can be contructed with different schemes of entanglement generation and CNOT gates. Below, we describe the constituents of these other schemes and compare them to those of the high-fidelity repeater

\subsection{Single-photon entanglement creation} \label{sec:1phot}
It has also been suggested to use single-photon detection schemes similar to Ref.~\cite{huelga} to generate entanglement in the elementary links. The setup of a single-photon detection scheme is also shown in \figref{fig:figure2}. We assume that the two emitters are initially prepared in a state 
\begin{equation}
(1-\epsilon^{2})\ket{00}+\epsilon^{2}\ket{ee}+\epsilon\sqrt{1-\epsilon^{2}}\left(\ket{0e}+\ket{e0}\right)
\end{equation}    
by a weak excitation pulse such that the excitation probability is $\epsilon^{2}$. An emitter can then go from state $\ket{e}$ to state $\ket{1}$ by emitting a cavity photon. The emitted photons are collected from the cavities and combined on a balanced beam splitter (BS) on a central station between the two cavities. Neglecting losses, the detection of a single photon after the BS will project the state of the emitters into the Bell state $\ket{\Psi^{+}}=\frac{1}{\sqrt{2}}\left(\ket{01}+\ket{10}\right)$ up to a single qubit rotation in the limit $\epsilon^{2}\ll1$ where we can neglect the possibility of double excitations. The probability of an emitter to go from $\ket{e}$ to $\ket{1}$, by emission of a cavity photon during a time interval $[0;T]$, is $P_{\text{phot}}$ (see Eq.~\eqref{eq:phot1}) under similar assumptions as in the two-photon scheme described above. Neglecting dark counts but including losses, the total probability of a single click at the central station is $P_{\text{1click}}=2\eta P_{\text{phot}}\epsilon^{2}(1-\epsilon^{2})+(2\eta-\eta^{2})P_{\text{phot}}^{2}\epsilon^{4}$ with $\eta$ being the total detection efficiency as for the two-photon scheme. The first term is the probability to emit and detect a single photon while the second term is the probability of emitting two cavity photons but only getting a single click (we assume that we do not have access to number-resolving detectors). The probability, to have a single click and have created the state $\ket{\Psi^{+}}$, is $P_{\text{correct}}=2\eta P_{\text{phot}}\epsilon^{2}(1-\epsilon^{2})$. The average heralded fidelity conditioned on a single click is thus $F_{1}=P_{\text{correct}}/P_{\text{1click}}$. To lowest order in $\epsilon$, $F_{1}\sim1-(1-\eta/2)P_{\text{phot}}\epsilon^{2}$ while the success probability is $P_{\text{1click}}\sim2\eta P_{\text{phot}}\epsilon^{2}$. There is thus a tradeoff set by $\epsilon^{2}$ between the success probability and the fidelity for the single-photon detection scheme. This is in contrast to the two-photon detection scheme where $F=1$ regardless of success probability. 

The success probability of the single-photon detection scheme is not as sensitive to the detection efficiency $\eta$ as the two-photon detection scheme as shown in \figref{fig:figureX3}.   
If the detection efficiency $\eta$ is large, the two-photon scheme is desirable since it will have both a high success probability and a high fidelity. However, if $\eta$ is small, the single-photon scheme will be advantageous since it has a relatively high success probability. Due to the possible high success probability but limited fidelity of the single-photon scheme, it might be desirable to combine it with entanglement purification to increase the final fidelity. In this way the higher success probability of the single-photon scheme may compensate the lower fidelity. We have therefore considered the possibility of initial entanglement purification in repeaters based on the single-photon detection scheme as described below.     

\subsection{Initial purification}
Based on a detailed analysis of the various errors that limit the fidelity for the single-photon scheme including dark counts of the detectors (see App.~\ref{single} for details), we find that the purification protocol of Ref.~\cite{bennett} effectively corrects for the errors in the single-photon scheme and we assume that this is used for the initial purification. However, as pointed out in Ref.~\cite{nickerson} an improved fidelity, at the expense of a factor of $\sim2$ in the success probability, can be obtained by only accepting outcomes where the two heralding qubits are found in state $\ket{1}\ket{1}$ instead of also accepting $\ket{0}\ket{0}$ outcomes. We will also consider this modification to the purification protocol in Ref.~\cite{bennett}. The protocol relies on a CNOT operation, which we assume to be made with the same gate used to perform the subsequent entanglement swapping (see below). To reflect the most realistic near-term quantum repeaters we consider at most 4 qubits per repeater stations. We therefore assume that the purification is performed in a pumping scheme~\cite{briegel2}, where the fidelity of a single pair is pumped by combining it with pairs of lower fidelity since this requires the lowest number of qubits per station. 

\begin{figure} 
\centering
\includegraphics[width=0.5\textwidth]{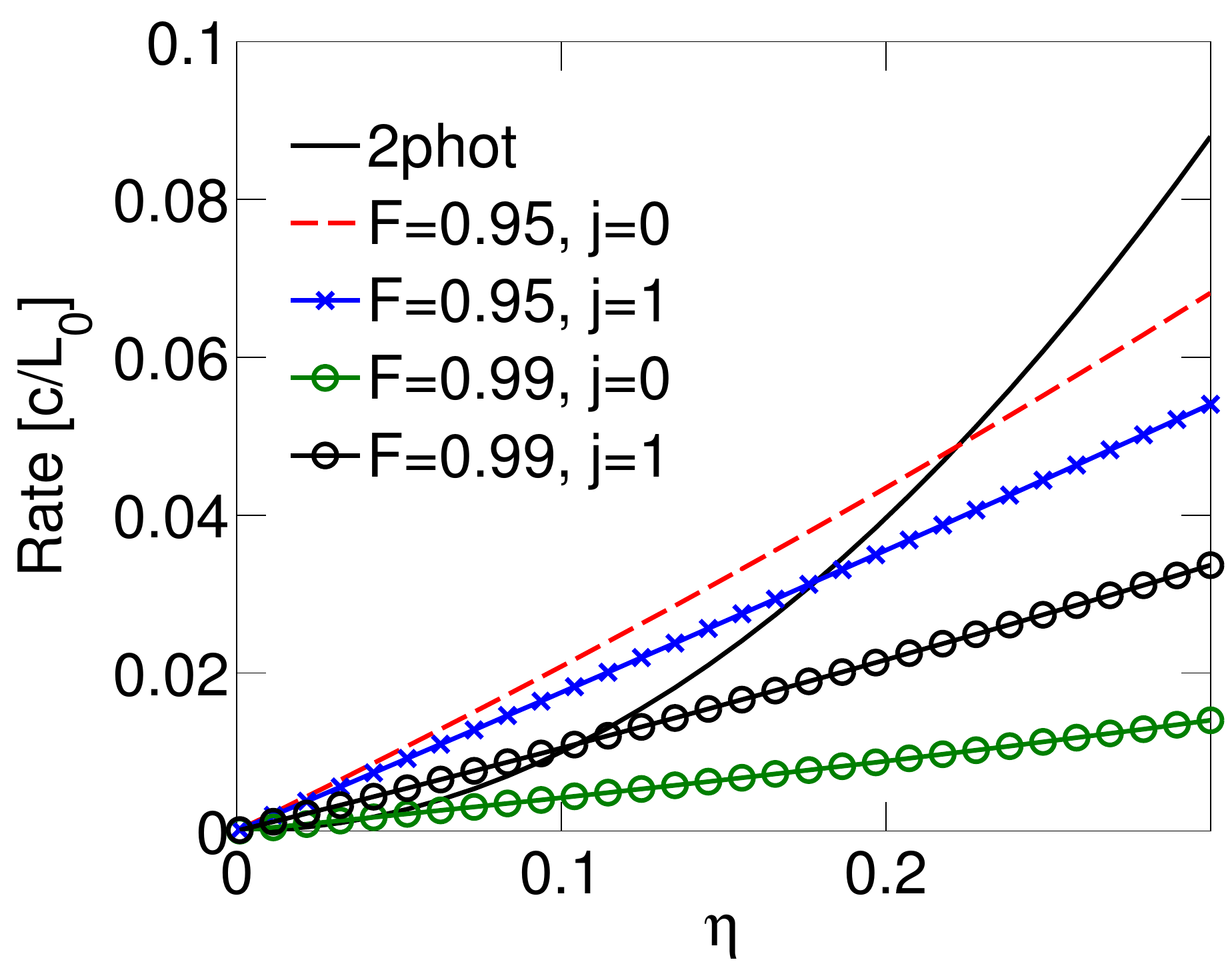}
\caption[Purification]{Rate of entanglement generation for the two-photon scheme and the single-photon scheme with target fidelity $F\geq0.95$ and $F\geq0.99$ both without purification ($j=0$) and with one round of purification ($j=1$). The rate is shown as a function of the total detection efficiency $\eta$. We have neglected dark counts and assumed that the CNOT gate is deterministic and have perfect fidelity. The rate has been calculated as described in App.~\ref{app:rate}. Furthermore, we have assumed that each repeater station contains 4 qubits, which are either used for purification or to increase the rate of the entanglement generation. }
\label{fig:figureX3}
\end{figure}

The effect of combining the single-photon scheme with initial purification is shown in \figref{fig:figureX3} where, for simplicity, the purification is assumed to be performed with a deterministic gate with perfect fidelity and without the modification of Ref.~\cite{nickerson}. If high fidelity pairs are desired for, e.g., a repeater with many swap levels, entanglement purification can increase the rate of the entanglement generation. For high collection efficiencies it is, however, desirable to use the two photon scheme since this has a higher rate. In particular the two photon scheme becomes desirable if high fidelity pairs are required.

\subsection{CNOT gates} \label{sec:CNOTgate}

\begin{figure}
\centering
\subfloat {\label{fig:figure4a}\includegraphics[width=0.25\textwidth]{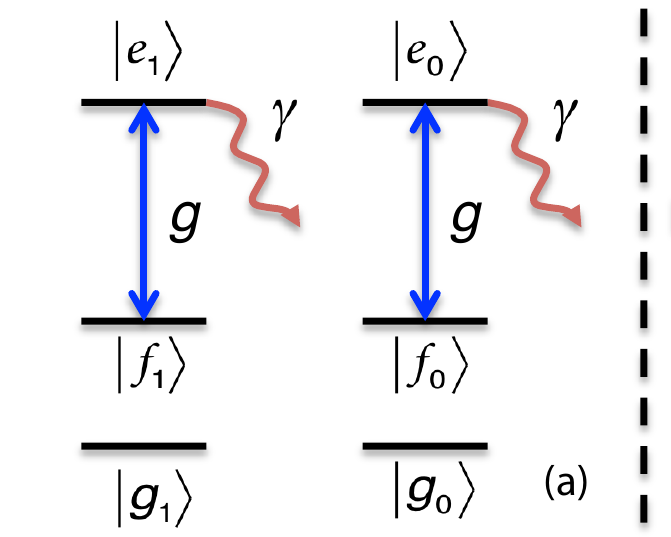}} 
\subfloat{\label{fig:figure4b}\includegraphics[width=0.25\textwidth]{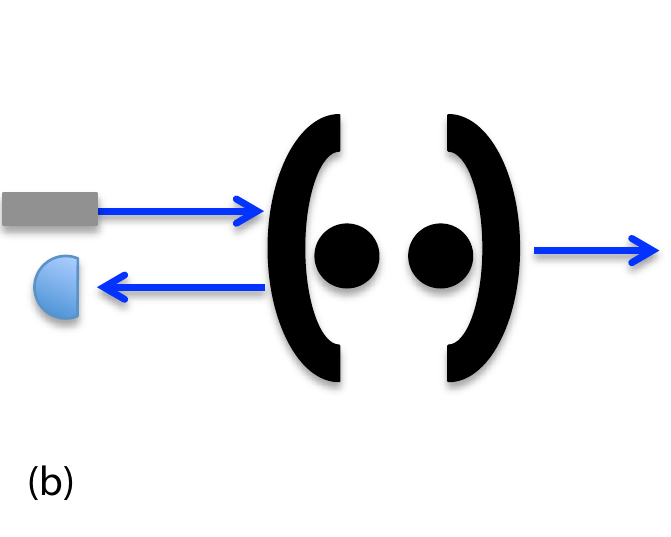}}
\caption[CNOT gate structure II]{Deterministic gate based on reflection and teleportation based CNOT operation~\cite{Anders1prl,Anders2prl}(a) Level structure of the qubit atoms. The levels $\ket{r_{0}}$ and $\ket{r_{1}}$ where $r=g,f$ or $e$ are assumed to be degenerate such that the quantum information is encoded in the horizontal degrees of freedom. (b) The setup to create entanglement between the level states $\ket{g},\ket{f}$ of the atoms. Weak coherent light is sent onto the cavity and any reflected light is measured with a SPD. }
\label{fig:figure4}
\end{figure}

In our analysis, both the initial purification and the subsequent entanglement swapping involves a cavity-based CNOT gate. Besides the heralded CNOT gate used in the high-fidelity repeater, which we will refer to as \emph{gate 1}, a deterministic cavity-based gate proposed in Ref.~\cite{Anders2prl} could be used. Combining the gate scheme of Ref.~\cite{Anders2prl} with the local entanglement generation scheme of Ref.~\cite{Anders1prl} results in a deterministic CNOT gate with an error scaling as $1/C$.  We will refer to this gate as \emph{gate 2}. This gate does not require an auxiliary atom as gate 1 but rather two auxiliary levels in the qubit atoms as shown in \figref{fig:figure4a}.
In this scheme, the quantum information is stored in the horizontal/qubit degree of freedom (subscripts $0$ and $1$) and the vertical/level degree of freedom (denoted $g$ and $f$) is used to make an entanglement assisted CNOT gate between the atoms. Separating the qubit degree of freedom from the level degree of freedom, the gate works by ideally making the transformation
\begin{equation} \label{eq:transgate1}
\ket{q1}\ket{q2}\otimes\ket{gg}\to\ket{q1}\ket{q2}\otimes\frac{1}{\sqrt{2}}\left(\ket{gf}+\ket{fg}\right),
\end{equation} 
where $\ket{g},\ket{f}$ denote the vertical states and $\ket{q1}$ ($\ket{q2}$) is the qubit state of the first (second) qubit, which could be entangled with atoms at neighboring repeater stations. The entanglement between the levels $\ket{g}$ and $\ket{f}$ can be used to make a CNOT gate, if the levels of the atoms can be measured non-destructively, i.e. without revealing any information about the qubit state as described in Ref.~\cite{Anders2prl}. Both the transformation shown in Eq.~\eqref{eq:transgate1} and the non-destructive measurements can be obtained by sending a weak coherent pulse onto a two-sided cavity and detecting any reflected light (see \figref{fig:figure4b} and App.~\ref{app:cnot}). If the light is resonant with the empty cavity mode, atoms in $\ket{f}$ will shift the cavity resonance. Consequently, photons will be reflected and constitutes a QND measurement of the presence of atoms in $\ket{f}$. Spontaneous emission from the atoms will limit the fidelity of the gate to $F\sim1-1.2/(\eta_{\text{d}}C)$, where $\eta_{\text{d}}$ is the detection efficiency and $C=g^{2}/\kappa\gamma$ is the cooperativity. The gate time is limited by the time of the single qubit rotations and the coherent pulses.  We assume that this gives a gate time on the order of 10 $\mu$s.  

As a benchmark, we also consider a naive approach where a direct gate between two qubits is made in a cavity without the use of an auxliary atom or auxliary atomic states. To characterize such a gate we consider a situation where the setup of gate 1 is used to make a deterministic gate by simply ignoring the heralding condition. We will refer to this gate as \emph{gate 3}. For such a gate, we find that the gate fidelity will scale as $1-F\sim3/\sqrt{C}$ and the time of the gate will be limited by the time of the single qubit rotations which we assume to be $\sim10$ $\mu$s. 

\begin{figure} 
\centering
\includegraphics[width=0.5\textwidth]{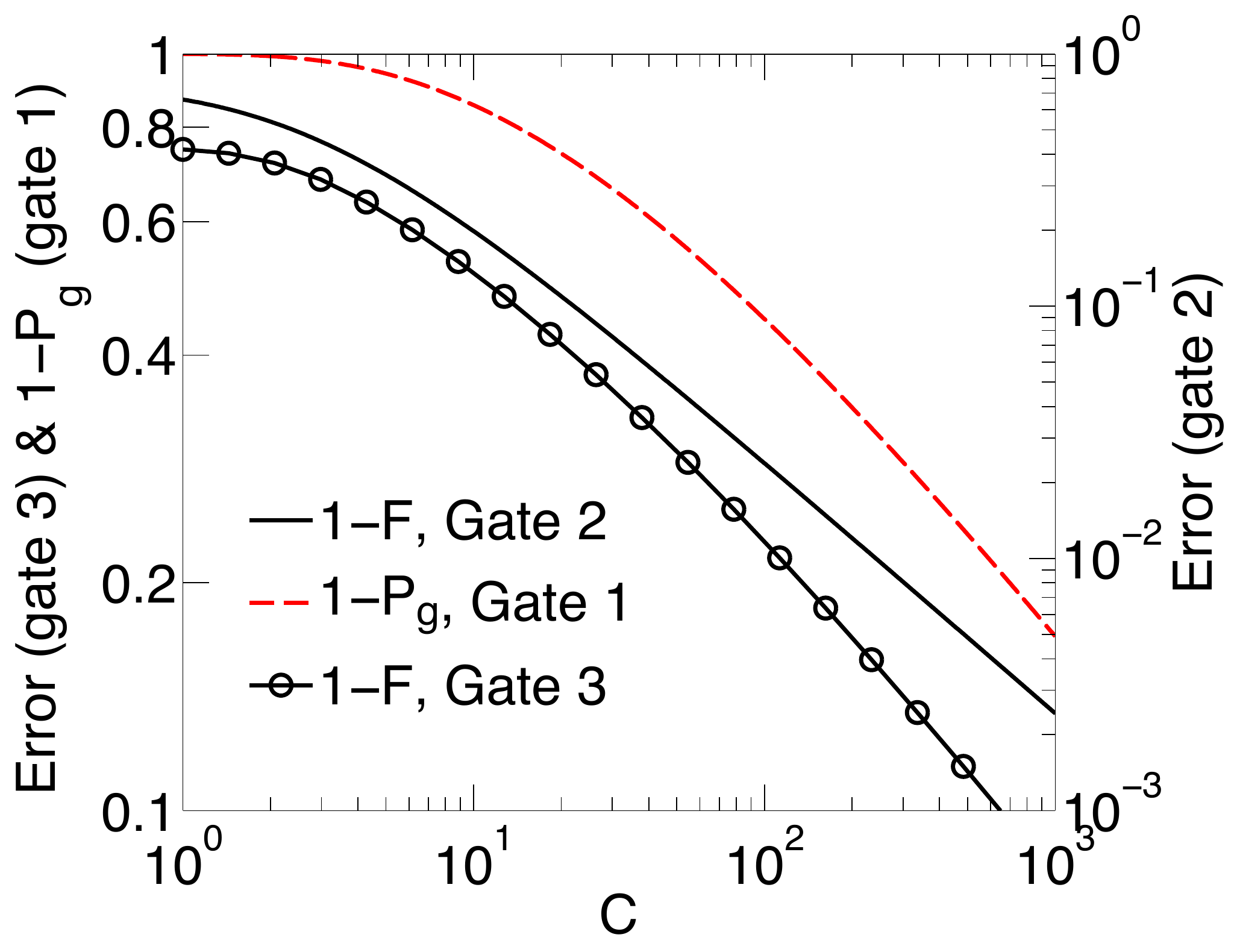}
\caption[CNOT gates comparison]{Characteristics of the three gates described in the text. The errors of gate 2 (black/solid line, right axis) and gate 3 (black/circled line, left axis) are shown as a function of the cooperativity. The error is defined as $1-F$ where $F$ is the fidelity of the gate. We have assumed $\eta_{d}=0.5$ for the error of gate 2. Gate 1 has conditional fidelity $\sim1$ but a finite failure probability $1-P_{g}$ which is also shown as a function of cooperativity (red/dashed line, left axis).}
\label{fig:figureX2}
\end{figure}

The characteristics of the three gates we consider are summarized in \tabref{tab:table2} and illustrated in \figref{fig:figureX2}.
\begin{table} 
\centering
\begin{tabular}{|c|c|c|c|}
\hline
Gate & Fidelity & Probability & Gate time  \\ \hline
1  & $F=4\cdot10^{-5}$ & $P_{g}\sim1-6/\sqrt{C}$ & $377/(\gamma\sqrt{C})$+10 $\mu$s\\ \hline
2 & $F\sim1-1.2/(\eta_{\text{d}}C)$ & $P_{g}=1$ &10 $\mu$s  \\ \hline
3 & $F\sim1-3.6/\sqrt{C}$ & $P_{g}=1$ &10 $\mu$s \\ \hline
\end{tabular}
\caption{The characteristics of the three gates considered for the cavity-based repeaters. $C$ is the cooperativity of the atom-cavity system and $\eta_{\text{d}}$ is the single photon detection efficiency in gate 2. The time of the single qubit rotations is assumed to be 10 $\mu$s. }
\label{tab:table2}
\end{table}
It is clear, that a repeater based on gate 3 will never be advantageous but we consider it as a reference since the physical requirements for implementing this gate are less than for gate 1 and 2, which requires either an auxiliary atom or auxiliary atomic levels.      

\section{Numerical optimization} \label{sec:optim}
\begin{figure} 
\centering
\includegraphics[width=0.46\textwidth]{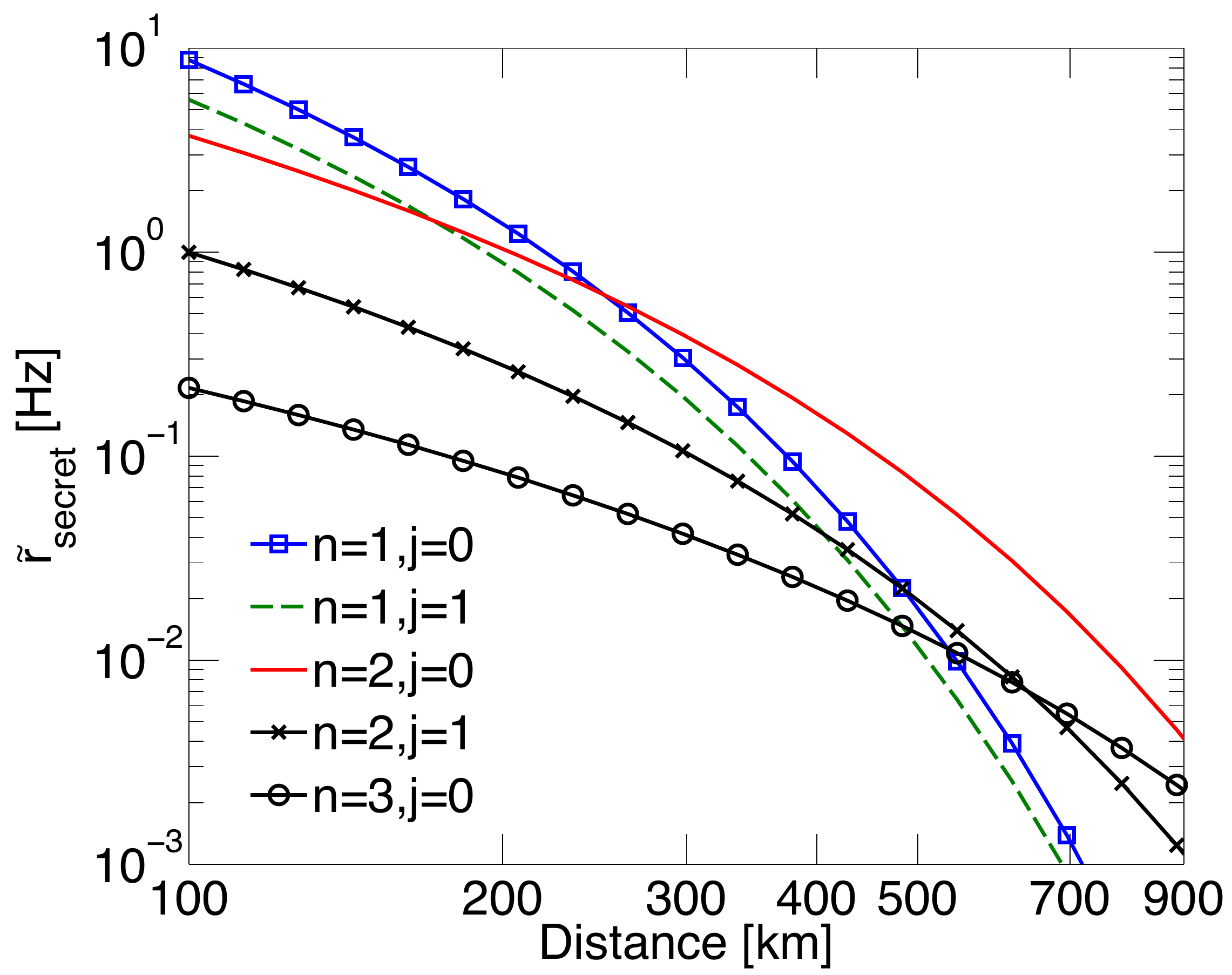}
\caption[Example of repeater architecture]{Normalized secret key rate per station($\tilde{r}_{\text{secret}}$) as a function of the distribution distance (Distance) for a parallel repeater based on the single photon generation scheme and gate 2. The cooperativity was assumed to be 100 and we assumed 4 qubits per repeater station. The optimal number of swap levels ($n$) and purification rounds ($j$) for a given distance can be directly read off from the plot as the combination giving the highest rate. Note that because the gate fidelity is limited, curves with $j=2$ and $n=3,j=1$ are not shown since they result in a much lower secret key rate. The purification schemes was considered to be without the modification of Ref.~\cite{nickerson}.}
\label{fig:figureX6}
\end{figure}

We have numerical optimized the secret key rate per repeater station for both the high-fidelity repeater and all other cavity-based repeaters consisting of the elements considered in Sec.~\ref{sec:other}. The secret key rate is calculated as described in Sec.~\ref{sec:secret} and App.~\ref{app:rate}. It depends on some experimental parameters such as the efficiency of single photon detectors, dark count rates etc. The values of these parameters are assumed fixed and are thus not part of the optimization.  All the experimental parameters are summarized in \tabref{tab:parameter} together with the values assumed in the optimizations. We have assumed fiber transmission losses for telecom wavelengths, which may require wavelength conversion techniques \cite{boris}.
\begin{table*}
\begin{tabular}{| c | c| p{8cm} | }
\hline
Parameter & Value & Description \\ \hline
$\gamma$ & $2\pi \cdot 6$ MHz & Spontaneous emission rate of atoms. This enters in the probability of emitting a photon in the entanglement generation schemes (see Eq.~\eqref{eq:phot1}) and in the gate time of gate 1 and 2. \\ \hline
$\eta_{\text{d}}$ & $50\%$ & Combined efficiency of SPD detectors and outcoupling of light from the cavities. This enters in the total detection efficiency $\eta$ in the entanglement generation schemes since $\eta=\eta_{\text{d}}\eta_{\text{f}}$. It also enters the fidelity of gate 2. \\ \hline 
$L_{att}$ & 22 km  & Attenuation length of the fibers. The total transmission probability over a length $L$ is assumed to be $\eta_{\text{f}}=e^{-L/L_{att}}$. The value assumed corresponds to telecom wavelengths.  \\ \hline
$\tau_{\text{local}}$ & 10 $\mu$s & Time of local qubit operations  \\ \hline
$r_{dark}$ & 25 Hz & Dark count rate of SPD detectors. We include dark counts in the entanglement generation step but not in the gate operations since the gate operations are assumed to be fast.  \\ \hline
$c$ & $2\cdot 10^{5}$ km/s & Reduced speed of light in the transmission fibers \cite{sangouard3}. \\ \hline
\end{tabular}
\caption[Parameters in the numerical optimization]{Experimental parameters which influence the rate and fidelity of the repeaters. The second column gives the values used in all optimizations.}
\label{tab:parameter}
\end{table*}
The free parameters in the optimizations are the number of swap levels, the number of purifications with/without the modification of Ref.~\cite{nickerson} and whether a parallel or sequential repeater protocol is used. In the optimizations, we calculate the secret key rate on a grid of all these parameters and pick the combination giving the highest rate. \figref{fig:figureX6} shows a specific example where the combination of the single-photon scheme with gate 2 is investigated for a parallel repeater and a cooperativity of 100. 
The number of swap levels and purifications, giving the highest rate for a specific distance, can be directly read off from the figure. The same calculations are then done for a sequential repeater protocol and compared to the parallel repeater protocol with/without the modified purification in order to find the highest rate for this specific combination of entanglement generation scheme and CNOT gate. This is done for all combinations of entanglement generation schemes and CNOT gates. The optimal evolution time, $T$, and excitation probability, $\epsilon$, in the entanglement generation schemes are found for each grid point using a built-in numerical optimization in the program MATLAB\footnote{see \href{http://www.mathworks.se/help/matlab/ref/fminsearch.html}{http://www.mathworks.se/help/matlab/ref/fminsearch.html}}. The key parameter, determining the performance of the CNOT gates, is the cooperativity (see \tabref{tab:table2}). We therefore optimize for cooperativities $C\in[10;1000]$ and distances between 100 km and 1000 km. Finally, the optimizations are performed for both 2 qubits per repeater station and 4 qubits per repeater station. Note that the auxiliary atom used in gate 1 is not counted as a qubit and schemes based on this gate thus in principle contain an additional atom per repeater station. 

We model the effect of the non-perfect gates, as depolarizing channels such that the output of a gate operation described by a unitary $U_{\mathcal{S}}$ working on a set $\mathcal{S}$ of two qubits is
\begin{equation} \label{eq:gateerror}
\tilde{\rho}=F'U_{\mathcal{S}}\rho U^{\dagger}_{\mathcal{S}}+\frac{1-F'}{4}\left(\text{Tr}\left\{\rho\right\}_{\mathcal{S}}\otimes\mathbb{1}_{S}\right), 
\end{equation}
where $F=F'+(1-F')/4$ is the fidelity of the gate, $\mathbb{1}_{\mathcal{S}}$ is the identity matrix of the set, $\text{Tr}\{\ldots\}_{\mathcal{S}}$ is the trace over the set and $\rho$ is the initial density matrix describing the system before the gate operation. We use Eq.~\eqref{eq:gateerror} to propagate the density matrix from the entanglement generation (see App.~\ref{single}-\ref{two}) through the steps of initial purification and entanglement swapping and calculate the average fidelity of the distributed pairs. To calculate the secret key fraction, we treat the distributed pairs as Werner states as described in \secref{sec:secret}.  
\begin{figure} [H]
\centering
\subfloat {\label{fig:figureX7a}\includegraphics[width=0.45\textwidth,height=2.7in]{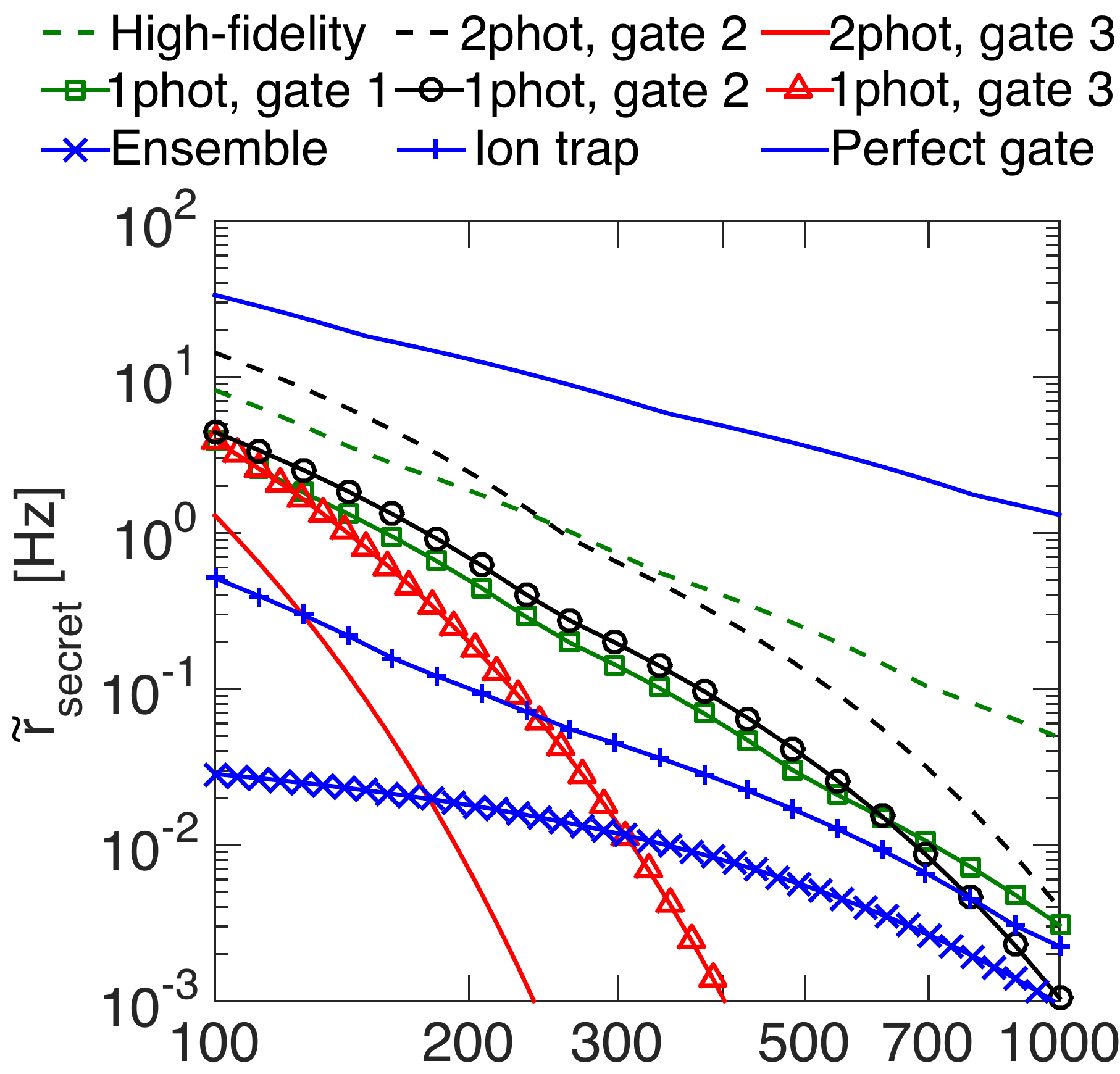}} \\
\subfloat{\label{fig:figureX7b}\includegraphics[width=0.45\textwidth,height=2.7in]{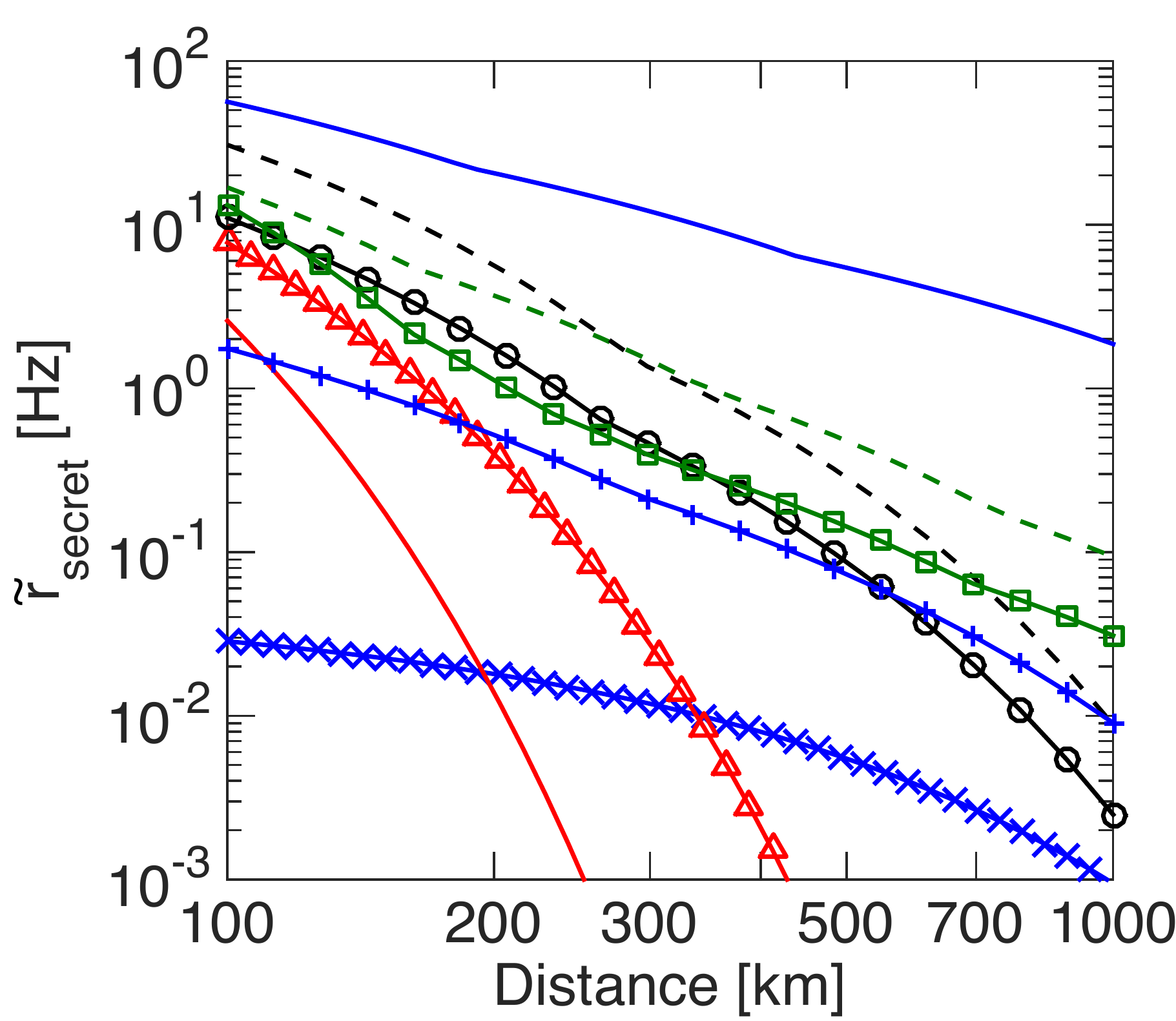}}
\caption[Optimal secret key rate I]{Normalized secret key rate per station ($\tilde{r}_{\text{secret}}$) as a function of the distribution distance (Distance) for the high-fidelity repeater and other cavity-based repeaters assuming a cooperativity of $C=$100. (a) is for 2 qubits per repeater station while (b) is for 4 qubits per repeater station. The other cavity-based repeaters are labelled as, e.g., ´``1phot, gate 2'', which indicates that it is a repeater based on the single-photon detection scheme and gate 2. The rate of an ensemble-based repeater (´``Ensemble'') is also shown~\cite{sangouard1}. For simplicity, we have assumed a fixed number of four swap levels in the ensemble-based repeater even though a smaller number of swap levels might increase the rate for small distances ($\lesssim400$ km). Finally, we have plotted the rate of an ion trap repeater scheme (´``Ion trap'') with a collection efficiency of 10\% and gate fidelity of 99.3\% and the ultimate rate obtainable with perfect deterministic gates and perfect entanglement generation with the two-photon scheme (´``Perfect gate'') for comparison. For the ion trap repeater we have plotted the highest rate obtainable with either the one-photon or two-photon scheme.}
\label{fig:figureX7}
\end{figure} 
\begin{figure} [H]
\centering
\subfloat {\label{fig:figureX8a}\includegraphics[width=0.45\textwidth, height=2.7in]{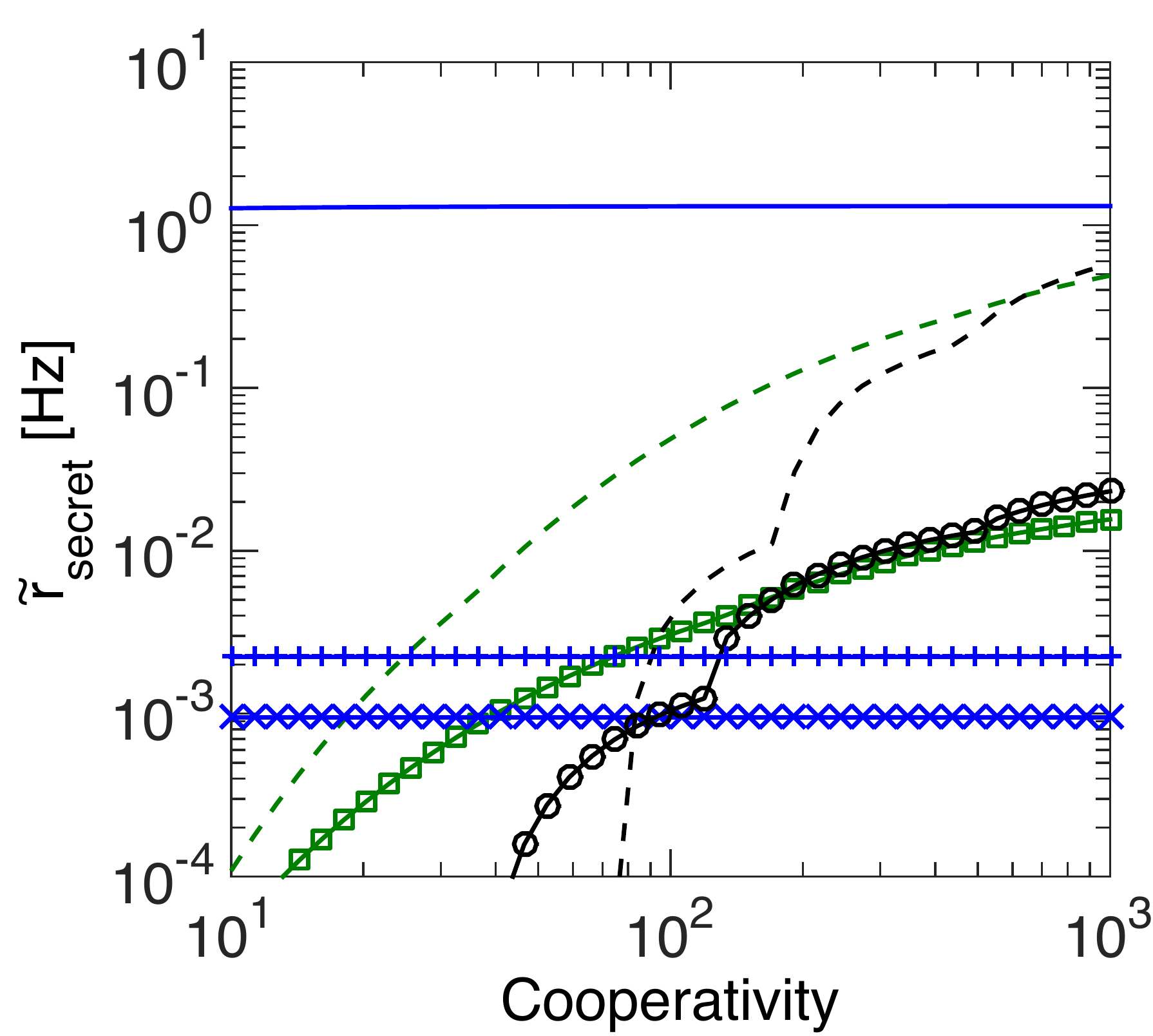}} \\
\subfloat{\label{fig:figureX8b}\includegraphics[width=0.45\textwidth,height=2.7in]{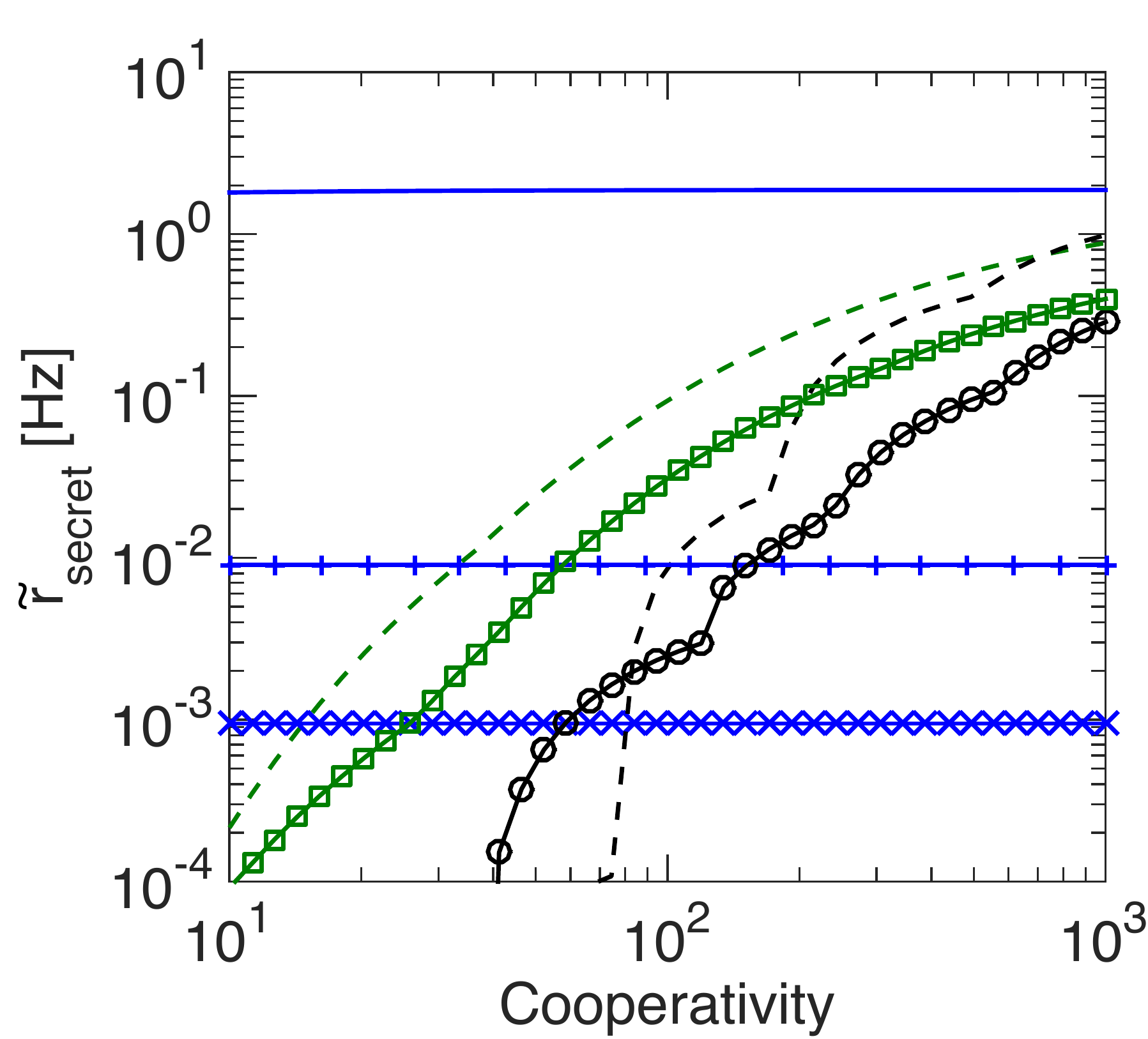}}
\caption[Optimal secret key rate II]{Normalized secret key rate per station ($\tilde{r}_{\text{secret}}$) as a function of the cooperativity ($C$) for the high-fidelity repeater and other repeaters assuming a distribution distance of 1000 km. (a) is for 2 qubits per repeater station while (b) is for 4 qubits per repeater station. The labels are the same as in \figref{fig:figureX7}}
\label{fig:figureX8}
\end{figure}

The secret key rates per repeater station of the high-fidelity repeater and the other cavity-based repeaters are shown in Fig.~\ref{fig:figureX7} for distances $[100;1000]$ km and a cooperativity of 100 and in Fig.~\ref{fig:figureX8} for a distance of 1000 km and cooperativities in the interval $[10;1000]$. As shown in \figref{fig:figureX7}, the repeaters based on gate 3 are simply not able to distribute entanglement over large distances for realistic cooperativities. As a consequence, repeaters based on gate 3 do not appear on Fig.~\ref{fig:figureX8} since their secret key rate is simply too low.

In general, the high-fidelity repeater (2-photon, gate 1) achieves the highest secret key rate for a broad range of cooperativities and long distances $\gtrsim 300$ km. This reflects both that this protocol allows for a higher number of swap levels and that the secret key rate favors the distribution of high-fidelity pairs since these gives the highest secret fraction (see \figref{fig:figure5}). It is also apparent from Fig.~\ref{fig:figureX8} that while repeaters based on gate 2 need cooperativities above 100 for a distance of 1000 km, repeaters based on gate 1 are able to function with much lower cooperativities around $30-40$. This is because the heralded gate has nearly unit fidelity independent of the cooperativity. For high cooperativities and/or low distances, a repeater based on the two-photon detection scheme and gate 2 can give a slightly higher secret key rate than the high-fidelity repeater. This improvement is, however, less than a factor of 2 in the secret key rate. The steps in the rates of the schemes based on gate 2 in Fig.~\ref{fig:figureX8} originate from the fact that as the cooperativity increases, the fidelity of gate 2 increases and at some point the fidelity is high enough to allow for another swap level, which makes the rate increase abruptly.  
From the optimizations, we find that the sequential repeater architecture achieves slightly higher rates (less than a factor 2) than the parallel repeater architecture for 4 qubits per repeater station, while the opposite is the case for 2 qubits per repeater station.   

In general, repeaters based on the two-photon detection schemes outperform repeaters based on the single-photon detection scheme except for repeaters based on gate 3. This reflects that repeaters based on gate 3 cannot perform many swap levels since the fidelity simply decreases too rapidly with the number of swap levels. The result of the optimization was that no swap levels were actually preferred for repeaters based on gate 3 for $C\leq1000$.  As a result, the elementary links in these repeaters are long and fiber losses therefore significantly decrease the total detection probability $\eta$ in the entanglement generation schemes. In the limit of very low $\eta$, the one-photon scheme is advantageous since the success probability only depend linearly on $\eta$. For the optimizations, we have assumed that the combined efficiency of the SPD detectors and outcoupling of light from the cavities is $\eta_{d}=$50\%. If this efficiency is smaller, repeaters based on single-photon detection may be desirable. We also find that the purification protocol, in general, performs better with than without the modification of Ref.~\cite{nickerson}. The improvement is, however, limited to a factor of $\lesssim2$ for the parameters considered in Figs.~\ref{fig:figureX7}-\ref{fig:figureX8}.   

It is important to stress that the rates plotted in Figs.~\ref{fig:figureX7}-\ref{fig:figureX8}, are the secret key rates divided by the total number of repeater stations. The actual distribution rate can thus be obtained by multiplying with the number of repeater stations. For the high-fidelity repeater, we find a secret key rate of $\sim$16 Hz over 1000 km for 33 repeater stations and a cooperativity of 1000 assuming 2 qubits per repeater station.  For a more modest cooperativity of 100, a secret key rate of $\sim$1.5 Hz over 1000 km can be obtained. 

We can compare the rate found here to the rate obtainable with repeaters based on atomic ensembles. In Ref.~\cite{sangouard1} an efficient repeater based on atomic ensembles is described, which achieves one of the highest distribution rates for repeaters based on atomic ensembles~\cite{sangouard3}. The fidelity of the distributed pair and the distribution rate are derived in Ref.~\cite{sangouard1} for a repeater with four swap levels corresponding to 17 repeater stations.  Based on this, we have calculated the secret key rate assuming an optimistic, basic repetition rate of the ensembles of 100 MHz and memory and SPD efficiencies of 90\%. The rate of the ensemble-based repeater is also shown in Figs.~\ref{fig:figureX7}-\ref{fig:figureX8} for similar assumptions about fiber losses etc. as for the cavity-based repeaters. We have assumed that the repeater uses four swap levels for all distances even though a smaller number of swap levels may be desirable for smaller distances ($\lesssim400$ km)~\cite{sangouard1}. For a distance of 1000 km, we find a rate of $\sim0.03$ Hz for 33 repeater stations. This shows that repeaters based on individual atoms in cavities may be very promising candidates for realizing efficient quantum repeaters with rates exceeding those obtainable with atomic ensembles. The main reason for this is that very efficient entanglement swapping can be realized in the cavity-based repeaters which greatly enhances the distribution rates for long distances. On the contrary, repeaters based on atomic ensembles and linear optics have an upper limit on the swapping efficiency of 50\%. 

For comparison we have also considered a repeater based on ion traps where there is no cavity to collect the light. Non-local entanglement can still be created by collecting the emitted light with a lens as demonstrated in Ref.~\cite{monroe2014} where a collection efficiency of 10\% was reported. The entanglement swapping can be realized using a gate, which has been demonstrated experimentally with a fidelity of 99.3\% and a gate time of $50$ $\mu$s~\cite{blatt1}. Note that this fidelity was measured for the generation of a single state in Ref.~\cite{blatt1} but we will assume it to be the fidelity of the entanglement swap. The rate of such a ion-trap repeater is shown in Figs.~\ref{fig:figureX7}-\ref{fig:figureX8} with assumptions about fiber losses etc. summarized in Tab.~\ref{tab:parameter}. We have assumed a collection efficiency of 10\% and as a result, the one photon scheme with modified purification performs better than the two-photon scheme for 4 qubits per repeater station. However for two qubits, where purification is not possible, the two-photon scheme is in general advantageous except for small distances ($<$ 200 km). We have assumed a gate fidelity of 99.3\% and have plotted the highest rate obtainable with either the one-photon or two-photon scheme.   

It is seen that the high-fidelity repeater outperforms the ion-trap repeater for $C\gtrsim30$, which is mainly due to the low collection efficiency in the entanglement generation. The ultimate rate obtainable with a repeater with perfect deterministic entanglement swapping and entanglement generation based on the two-photon detection scheme with a collection efficiency set by $4C/(1+4C)$ is also shown in Figs.~\ref{fig:figureX7}-\ref{fig:figureX8} under our assumptions about fiber loss, detection efficiency etc.. A similar repeater was considered in Ref.~\cite{sangouard2} to demonstrate the feasibility of repeaters based on trapped ions. For $C=1000$, the high-fidelity repeater achieves only a factor of $\sim2$ slower rate than this ultimate limit for a distance of 1000 km. 

\section{Conclusion}        

In conclusion, we have performed a detailed analysis of quantum repeaters based on individual emitters in optical cavities. We have found that a high-fidelity repeater based on the heralded gate described in Ref.~\cite{johannes} combined with a two-photon detection scheme is the best option over a large parameter regime and enables high secret key rates over large distances even for limited cooperativities $<100$. Compared with a number of other cavity based repeaters it achieves rates, which are up to two orders of magnitude faster for long distances (1000 km) and cooperativities $<100$. For small distances or higher cooperativities, a repeater based on the deterministic CNOT gate described in Ref.~\cite{Anders2prl} combined with a two-photon detection scheme can achieve rates which are slightly higher than the high-fidelity repeater but the improvement is less than a factor of 2. 

We have also compared the high-fidelity repeater to the repeater in Ref.~\cite{sangouard1}, which is based on atomic ensembles. For a distance of 1000 km and $C\gtrsim20$ the high-fidelity repeater begins to outperform the ensemble-based repeater and an improvement of more than two orders of magnitude in the secret key rate is possible for $C\gtrsim100$. The main reason for the advantage of the high-fidelity repeater is that entanglement can be swapped very efficiently using the heralded CNOT gate described in Ref.~\cite{johannes}. Consequently, the number of swap levels in the repeater can be increased without the need of intermediate purification which greatly enhances the rate for large distances. A similar advantage could in principle be achieved by resorting to a trapped ion system where efficient gates can be implemented. For current systems, the collection efficiencies are, however, so low that a trapped ion system could be outperformed by a cavity system with a limited finesse of $C\gtrsim 30$. If the collection efficiency could be overcome, e.g. by placing the ions in a cavity with a high cooperativity, the rate can be substantially improved, but with $C>1000$ the high fidelity repeater investigated here is within a factor of two of this ideal repeater. It should, however, be noted that we have compared schemes with strong physical differences in our analysis. The high-fidelity repeater requires an extra auxiliary atom while auxiliary atomic levels are required to decrease the error of the deterministic cavity-based CNOT gates. The ensemble-based and ion-trap repeaters are also very different physical systems compared to the cavity-based repeaters with individual atoms. The different experimental difficulties in realizing the physical requirements for the various repeater schemes should be included in a more advanced assessment. 

Finally, we note that while we have investigated a number of different possible repeater protocols there may be even more advantages procedures. Hence the results that we have derived here should be seen as lower limits to the achievable communication rates. A particular interesting  possibility could be to investigate proposals along the lines of Ref.~\cite{cirac1,cirac2}, which also rely on heralding measurements to detect errors during entanglement generation and two qubit operations.  Possibly some of the ideas from these schemes could be used to improve the communication beyond what we have found here.

\begin{acknowledgements}
We gratefully acknowledge the support from the Lundbeck Foundation, the Carlsberg Foundation, NSF, CUA, DARPA, AFOSR MURI, and ARL. The research leading to these results has also received funding from the European Research Council under the European Union’s Seventh Framework Programme (FP/2007-2003) through SIQS (Grant No. 600645) and ERC Grant QIOS (Grant No. 306576). 
\end{acknowledgements}  

\appendix

\section{Error analysis of the single-photon scheme} \label{single}

The setup of the single-photon scheme is described in Sec. \ref{sec:1phot}. The single photon detectors are assumed to have a dark count probability of $P_{\text{dark}}$ and an efficiency of $\eta_{\text{d}}$ while the transmission efficiency of the fibers is denoted $\eta_{\text{f}}$.  As described in Sec. \ref{sec:generation}, the probability of an emitter to go from the excited state, $\ket{e}$ to the ground state $\ket{1}$ is $P_{\text{phot}}$ while the excitation probability is $\epsilon^{2}$. The scheme is conditioned on a single click at the central station. Depending on which detector gave the click, a single qubit rotation can be employed such that ideally the state $\ket{\Psi^{+}}$ is created. Going through all the possibilities of obtaining a single click at the central station, we find that the density matrix following a single click, and possible subsequent single qubit rotations, is
\begin{eqnarray}
\rho_{1click}&=&F_{1}\ket{\Psi^{+}}\bra{\Psi^{+}}+\alpha_{1}\ket{\Phi^{+}}\bra{\Phi^{+}}+\alpha_{1}\ket{\Phi^{-}}\bra{\Phi^{-}}\qquad\nonumber \\
&&+\beta_{1}\ket{\Psi^{-}}\bra{\Psi^{-}}+\tilde{\alpha}_{1}\ket{00}\bra{00}+\tilde{\beta}_{1}\ket{11}\bra{11},
\end{eqnarray}
with coefficients
\begin{eqnarray}
F_{1}&=&\frac{1}{P_{\text{1click}}}\Big[2\eta_{\text{d}}\eta_{\text{f}}P_{\text{phot}}\epsilon^{2}(1-\epsilon^{2})(1-P_{\text{dark}})\nonumber \\
&&+2\eta_{\text{f}}(1-\eta_{\text{d}})P_{\text{phot}}\epsilon^{2}(1-\epsilon^{2})P_{\text{dark}}(1-P_{\text{dark}})\nonumber \\
&&+\frac{1}{2}\eta_{\text{d}}\eta_{\text{f}}P_{\text{phot}}\epsilon^{4}(1-P_{\text{phot}}(1-P_{\text{dark}})+\nonumber \\
&&+2(1-\eta_{\text{f}})P_{\text{phot}}\epsilon^{2}(1-\epsilon^{2})P_{\text{dark}}(1-P_{\text{dark}}) \nonumber \\
&&+(1-\eta_{\text{d}}\eta_{\text{f}})P_{\text{phot}}\epsilon^{4}(1-P_{\text{phot}})P_{\text{dark}}(1-P_{\text{dark}})\nonumber \\
&&+(1-\epsilon^{2})\epsilon^{2}(1-P_{\text{phot}})P_{\text{dark}}(1-P_{\text{dark}})\nonumber \\
&&+\frac{1}{2}\epsilon^{2}(1-P_{\text{phot}})^{2}P_{\text{dark}}(1-P_{\text{dark}})\Big] \nonumber \\
\alpha_{1}&=&\frac{1}{P_{\text{1click}}}\Big[\frac{1}{2}\epsilon^{2}(1-P_{\text{phot}})^{2}P_{\text{dark}}(1-P_{\text{dark}})\Big] \nonumber \\
\beta_{1}&=&\frac{1}{P_{\text{1click}}}\Big[\frac{1}{2}\eta_{\text{d}}\eta_{\text{f}}P_{\text{phot}}\epsilon^{4}(1-P_{\text{phot}})(1-P_{\text{dark}})\nonumber \\
&&+2(1-\eta_{\text{f}})P_{\text{phot}}\epsilon^{2}(1-\epsilon^{2})P_{\text{dark}}(1-P_{\text{dark}}) \nonumber \\
&&+(1-\eta_{\text{d}}\eta_{\text{f}})P_{\text{phot}}\epsilon^{4}(1-P_{\text{phot}})P_{\text{dark}}(1-P_{\text{dark}})\nonumber \\
&&+(1-\epsilon^{2})\epsilon^{2}(1-P_{\text{phot}})P_{\text{dark}}(1-P_{\text{dark}}) \nonumber \\
&&+\frac{1}{2}\epsilon^{2}(1-P_{\text{phot}})^{2}P_{\text{dark}}(1-P_{\text{dark}}) \nonumber \\
&&+2\eta_{\text{f}}(1-\eta_{\text{d}})P_{\text{phot}}\epsilon^{2}(1-\epsilon^{2})P_{\text{dark}}(1-P_{\text{dark}})\Big] \nonumber \\
\tilde{\alpha}_{1}&=&\frac{1}{P_{\text{1click}}}\Big[2(1-\epsilon^{2})^{2}P_{\text{dark}}(1-P_{\text{dark}})\nonumber \\
&&+2(1-\epsilon^{2})\epsilon^{2}(1-P_{\text{phot}})P_{\text{dark}}(1-P_{\text{dark}})\Big] \nonumber \\
\tilde{\beta}_{1}&=&\frac{1}{P_{\text{1click}}}\Big[\eta_{\text{d}}\eta_{\text{f}}P_{\text{phot}}\epsilon^{4}(1-P_{\text{phot}})(1-P_{\text{dark}})\nonumber \\
&&+2(1-\eta_{\text{d}}\eta_{\text{f}})P_{\text{phot}}\epsilon^{4}(1-P_{\text{phot}})P_{\text{dark}}(1-P_{\text{dark}}) \nonumber \\
&&+2(1-\eta_{\text{d}}\eta_{\text{f}})^{2}P_{\text{phot}}^{2}\epsilon^{4}P_{\text{dark}}(1-P_{\text{dark}})\nonumber \\
&&+2(1-\eta_{\text{d}}\eta_{\text{f}})\eta_{\text{d}}\eta_{\text{f}}P_{\text{phot}}^{2}\epsilon^{4}(1-P_{\text{dark}})^{2}\Big]. 
\end{eqnarray}
Here we have assumed that with probability $\epsilon^{2}(1-P_{\text{phot}})$, an emitter is excited but spontaneously decay to the ground states instead of emitting a cavity photon. Furthermore, we have assumed that the decay rates to the two ground states are equal such that the emitter ends up in $\frac{1}{2}(\ket{0}\bra{0}+\ket{1}\bra{1})$. Note that the detectors are not assumed to be number resolving.  $P_{\text{1click}}$ is the total success probability given by
\begin{eqnarray}
P_{\text{1click}}&=&2\eta_{\text{d}}\eta_{\text{f}}P_{\text{phot}}\epsilon^{2}(1-P_{\text{phot}}\epsilon^{2})(1-P_{\text{dark}})\nonumber \\
&&+(2\eta_{\text{f}}\eta_{\text{d}}-\eta_{\text{f}}^{2}\eta_{\text{d}}^{2})P_{\text{phot}}^{2}\epsilon^{4} \nonumber \\
&&+2(1-\epsilon^{2}P_{\text{phot}})^{2}P_{\text{dark}}(1-P_{\text{dark}})\nonumber \\
&&+2(1-\eta_{\text{d}}\eta_{\text{f}})^{2}P_{\text{phot}}^{2}\epsilon^{4}P_{\text{dark}}(1-P_{\text{dark}})\nonumber \\
&&+4(1\!-\!\eta_{\text{d}}\eta_{\text{f}})P_{\text{phot}}\epsilon^{2}\nonumber \\
&&\times(1\!-\!\epsilon^{2}P_{\text{phot}})P_{\text{dark}}(1\!-\!P_{\text{dark}}).
\end{eqnarray}
Assuming $P_{\text{dark}}\ll1$, the dominant error is where both qubits are excited but only a single click is detected at the central station. This leaves the qubits in the state $\ket{11}\bra{11}$ and this error is efficiently detected by the purification scheme described in Ref. \cite{bennett}. 

\section{Error analysis of the two-photon scheme} \label{two}

For the two photon scheme described in Sec. \ref{sec:generation}, we condition on a click in two detectors. Once again we assume that appropriate single qubit rotations are employed depending on which detector combination clicked such that ideally the state $\ket{\Phi^{+}}$ is created. We find that the density matrix describing the qubit state after a successful event is 
\begin{eqnarray}
\rho_{2click}&=&F_{2}\ket{\Phi^{+}}\bra{\Phi^{+}}+\alpha_{2}\ket{\Psi^{+}}\bra{\Psi^{+}}\nonumber \\
&&+\alpha_{2}\ket{\Psi^{-}}\bra{\Psi^{-}}+\beta_{2}\ket{\Phi^{-}}\bra{\Phi^{-}},
\end{eqnarray} 
where we have defined
\begin{eqnarray}
F_{2}&=&\frac{(1-P_{\text{dark}})^{2}}{P_{\text{2click}}}\Big[\frac{1}{2}\eta_{d}^{2}\eta_{\text{f}}^{2}P_{\text{phot}}^{2} \nonumber \\
&&+\eta_{\text{d}}(1-\eta_{\text{d}})\eta_{\text{f}}^{2}P_{\text{dark}}P_{\text{phot}}^{2}\nonumber \\
&&+\eta_{\text{f}}^{2}(1-\eta_{\text{d}})^{2}P_{\text{phot}}^{2}P_{\text{dark}}^{2}\nonumber \\
&&+P_{\text{dark}}^{2}(1-P_{\text{phot}})^{2}+\eta_{\text{d}}(1-\eta_{\text{f}})\eta_{\text{f}}P_{\text{dark}}P_{\text{phot}}^{2}\nonumber \\
&&+\eta_{\text{d}}\eta_{\text{f}}P_{\text{dark}}P_{\text{phot}}(1-P_{\text{phot}})\nonumber \\
&&+(1-\eta_{\text{f}})^{2}P_{\text{phot}}^{2}P_{\text{dark}}^{2}\nonumber \\
&&+2\eta_{\text{f}}(1-\eta_{\text{d}})(1-\eta_{\text{f}})P_{\text{phot}}^{2}P_{\text{dark}}^{2}\nonumber \\
&&+2(1-\eta_{\text{d}}\eta_{\text{f}})P_{\text{phot}}(1-P_{\text{phot}})P_{\text{dark}}^{2}\Big] \nonumber \\
\alpha_{2}&=&\frac{(1-P_{\text{dark}})^{2}}{P_{\text{2click}}}\Big[\eta_{d}(1-\eta_{\text{f}})\eta_{\text{f}}P_{\text{dark}}P_{\text{phot}}^{2}\nonumber \\
&&+P_{\text{dark}}^{2}(1-P_{\text{phot}})^{2}+\eta_{\text{d}}\eta_{\text{f}}P_{\text{dark}}P_{\text{phot}}(1-P_{\text{phot}})\nonumber \\
&&+(1-\eta_{\text{f}})^{2}P_{\text{phot}}^{2}P_{\text{dark}}^{2}\nonumber \\
&&+2\eta_{\text{f}}(1-\eta_{\text{d}})(1-\eta_{\text{f}})P_{\text{phot}}^{2}P_{\text{dark}}^{2}\nonumber \\
&&+2(1-\eta_{\text{d}}\eta_{\text{f}})P_{\text{phot}}(1-P_{\text{phot}})P_{\text{dark}}^{2}\nonumber \\
&&+\eta_{\text{d}}(1-\eta_{\text{d}})\eta_{\text{f}}^{2}P_{\text{dark}}P_{\text{phot}}^{2}\nonumber \\
&&+\eta_{\text{f}}^{2}(1-\eta_{\text{d}})^{2}P_{\text{phot}}^{2}P_{\text{dark}}^{2}\Big] \nonumber \\
\beta_{2}&=&\alpha_{2}+\frac{(1-P_{\text{dark}})^{2}}{P_{\text{2click}}}\Big[\eta_{\text{d}}(1-\eta_{\text{d}})\eta_{\text{f}}^{2}P_{\text{dark}}P_{\text{phot}}^{2}\nonumber \\
&&+\eta_{\text{f}}^{2}(1-\eta_{\text{d}})^{2}P_{\text{phot}}^{2}P_{\text{dark}}^{2}\Big]. 
\end{eqnarray}
The success probability $P_{\text{2click}}$ is 
\begin{eqnarray}
P_{\text{2click}}&=&(1-P_{\text{dark}})^{2}\Big[\frac{1}{2}\eta_{\text{d}}^{2}\eta_{\text{f}}^{2}P_{\text{phot}}^{2} \nonumber \\
&&+4\eta_{\text{d}}\eta_{\text{f}}(1-\eta_{\text{d}}\eta_{\text{f}})P_{\text{dark}}P_{\text{phot}}^{2}\nonumber \\
&&+4P_{\text{dark}}^{2}(1-P_{\text{dark}})^{2} \nonumber \\
&&+4\eta_{\text{d}}\eta_{\text{f}}P_{\text{dark}}P_{\text{phot}}(1-P_{\text{phot}})\nonumber \\
&&+4(1-\eta_{\text{d}}\eta_{\text{f}})^{2}P_{\text{phot}}^{2}P_{\text{dark}}^{2} \nonumber \\
&&+8(1-\eta_{\text{d}}\eta_{\text{f}})P_{\text{phot}}(1-P_{\text{phot}})P_{\text{dark}}^{2}\Big]
\end{eqnarray}
As in the single-photon scheme, we have not assumed number resolving detectors and we have assumed that with probability ($1-P_{\text{phot}}$), an emitter spontaneously decay to one of the ground states resulting in the state $\frac{1}{2}(\ket{0}\bra{0}+\ket{1}\bra{1})$. 

\section{Deterministic CNOT gate} \label{app:cnot}

Here we describe the local entanglement generation scheme presented in Ref.~\cite{Anders1prl}, which can be used to make a deterministic CNOT gate as described in \secref{sec:CNOTgate}. We assume that weak coherent light is continuously shined onto the cavity such that at most one photon is in the cavity at all times. A single-photon detector continuously monitors if any photons are reflected from the cavity and the coherent light is blocked if a click is recorded before $n_{max}$ photons on average have been sent onto the cavity. If no click was recorded during this time, both atoms are interpreted as being in the $g$ levels. The steps of the entangling scheme are the following.
\begin{enumerate}
\item  Both atoms are initially prepared in the superposition $\ket{g}+\ket{f}$ by e.g. a $\pi/2$-pulse. 
\item Coherent light is sent onto the cavity.  If a click is recorded before on average $n_{\text{max}}$ photons have been sent onto the cavity, the levels of the atoms are flipped ($\ket{g}\leftrightarrow\ket{f}$). If no click is recorded, the atoms are interpreted as being in $\ket{gg}$ and the procedure is repeated from step 1. 
\item Conditioned on the first click, another coherent light pulse is sent onto the cavity after the levels of the atoms have been flipped. If a click is recorded before $n=n_{\text{max}}-n_{1}$ photons on average have been sent onto the cavity, the entangling scheme is considered to be a success. Here $n_{1}$ is the average number of photons that had been sent onto the cavity before the first click. If no click is recorded, the atoms are interpreted as being in $\ket{gg}$ and the procedure is repeated from step 1.   
\end{enumerate}
As described above, the entangling scheme is repeated until it is successful leading to a deterministic creation of entanglement in the end. As described in Ref. \cite{Anders2prl} a series of non-destructive measurements of the atoms together with single qubit rotations can be used to make a CNOT operation after the entanglement has been created. The non-destructive measurements can be performed using the same technique of monitoring reflected light as in the entangling scheme and we assume that we can effectively tune the couplings to the cavity such that possibly only a single atom couples.

\section{Rate analysis} \label{app:rate}

Here we analyse the rate of entanglement distribution for the different repeater architectures considered in the main text. The total rate of the repeater is set by the average time of entanglement creation, initial purification and entanglement swapping. Assuming that entanglement generation has a success probability, $P_{0}$, we estimate the average time $\tau_{\text{pair},l;m}$ it takes to generate $l$ entangled pairs in one elementary link using $m$ qubits, which can be operated in parallel, as
\begin{equation}
\tau_{\text{pair},l;m}=\mathcal{Z}_{l;m}(P_{0})(L_{0}/c+\tau_{\text{local}}).
\end{equation}
Here $c$ is the speed of light in the fibers and $\tau_{\text{local}}$ is the time of local operations such as initialization of the qubits. The factor $\mathcal{Z}_{l;m}(P_{0})$ can be thought of as the average number of coin tosses needed to get at least $l$ tails if we have access to $m$ coins, which we can flip simultaneously and the probability of tail is $P_{0}$ for each coin \cite{bernardes}. It is furthermore assumed that coins showing tail after a toss are kept and only the coins showing head are tossed again until $l$ tails are obtained. In the repeater context, the coins are entanglement generation attempts and tail is successful entanglement generation. The time it takes per "toss" is $L_{0}/c+\tau_{\text{local}}$. To calculate the expressions for $Z_{l;m}(P_{0})$, we follow the lines of Ref. \cite{bernardes} where similar factors are derived. The expression for $\mathcal{Z}_{m;m}(p)$ is already derived in Ref. \cite{bernardes} and their result is stated below
\begin{equation}
\mathcal{Z}_{m;m}=\sum_{k=1}^{m}\binom{m}{k}\frac{(-1)^{k+1}}{1-(1-p)^{k}}. 
\end{equation}
For $\mathcal{Z}_{l;m}$ where $l\neq m$, we only need to find expressions for $\mathcal{Z}_{1;m}$ with $m=1,2,3,4$, $\mathcal{Z}_{2;m}$ with $m=3,4$ and $\mathcal{Z}_{3;4}$ since we have a maximum of 4 qubits pr. repeater station. For $\mathcal{Z}_{2;3}$, we have that
\begin{eqnarray} \label{eq:suppf}
\mathcal{Z}_{2:3}&=&\binom{3}{3}\sum_{k=1}^{\infty}k(q^{3})^{k-1}p^{3}+\binom{3}{2}\sum_{k=1}^{\infty}k(q^{3})^{k-1}p^{2}q \qquad \nonumber \\
&&+\binom{3}{1}\binom{2}{1}\sum_{k=1}^{\infty}\sum_{l=1}^{\infty}(k+l)[(q^{3})^{k-1}pq^{2}][(q^{2})^{l-1}pq] \nonumber \\
&&+\binom{3}{1}\binom{2}{2}\sum_{k=1}^{\infty}\sum_{l=1}^{\infty}(k+l)[(q^{3})^{k-1}pq^{2}]\nonumber \\
&&\times[(q^{2})^{l-1}p^{2}],
\end{eqnarray}
where $q=1-p$. 
The first term in Eq.~\eqref{eq:suppf} describes the situations where three tails are obtained in a single toss after a given number of tosses, where all coins showed head.  We will refer tosses where all coins show tail as failed tosses. The second term describes the situation where we get two tails in the same toss after a given number of failed tosses. The third and fourth terms are where we get a single tail after a given number of failed tosses. The coin showing tail is then kept and the two remaining coins are tossed until we obtain another tail (third term) or two tails simultaneously (fourth term). The geometric series in Eq.~\eqref{eq:suppf} can be solved to give
\begin{equation}
\mathcal{Z}_{2;3}=\frac{5-(7-3p)}{(2-p)p(3+(p-3)p)} \approx\frac{5}{6p},
\end{equation} 
where the approximate expression is for $p\ll1$. Note that the factor of $\frac{5}{6}$ corresponds to a simple picture where it on average takes $\frac{1}{3}\frac{1}{p}$ attempts to get the first 'tail' using 3 coins and $\frac{1}{2}\frac{1}{p}$ attempts to get the second using the remaining $2$ coins. In a similar manner, we find that
\begin{eqnarray}
\mathcal{Z}_{1;2}&=&\frac{1}{2p-p^{2}}\approx\frac{1}{2p} \\
\mathcal{Z}_{1;3}&=&\frac{1}{3p-3p^{2}+p^{3}}\approx\frac{1}{3p} \\
\mathcal{Z}_{1;4}&=&\frac{1}{4p-6p^{2}+4p^{3}-p^{4}}\approx\frac{1}{4p} \\
\mathcal{Z}_{2;4}&=&\frac{-7+p(15+p(4p-13))}{(p-2)p(3+(p-3)p)(2+(p-2)p)}\nonumber \\
&\approx&\frac{7}{12p} \\
\mathcal{Z}_{3;4}&=&\frac{-13+p(33+p(22p-6p^{2}-37))}{(p-2)p(3+(p-3)p)(2+(p-2)p)}\nonumber \\
&\approx&\frac{13}{12p}. 
\end{eqnarray}
Here the approximate expressions are all for $p\ll1$ and they correspond to the expressions one would get using simple pictures similar to the one described above in the discussion of $\mathcal{Z}_{2;3}$. 

After creating a number of entangled pairs in an elementary link of the repeater, they may be combined to create a purified pair of higher fidelity. As previously mentioned, we assume an entanglement pumping scheme since this requires less qubit resources than a cascading scheme. Let $P_{\text{pur}}(F_{0},F_{0})$ denote the success probability of the purification operation, which depends on the fidelity of the two initial pairs ($F_{0}$) and the fidelity of the CNOT gate used in the purification operation. Note that $P_{\text{pur}}$ also contains the success probability of the CNOT gate used in the purification for the heralded gate. We estimate the average time $\tau_{\text{pur},1}$, it takes to make one purified pair from two initial pairs of fidelity $F_{0}$, using $m$ qubits in parallel in the entanglement generation step, as
\begin{equation}
\tau_{\text{pur},1}=\frac{\tau_{\text{pair},2;m}+\tau_{\text{pur}}}{P_{\text{pur}}(F_{0},F_{0})},
\end{equation}     
where $\tau_{\text{pur}}\sim L_{0}/c+\tau_{\text{c}}$ is the time of the purification operation including the classical comunication time required to compare results. Here $\tau_{\text{c}}$ is the time of the CNOT operation and $L_{0}/c$  is the communication time between the two repeater stations sharing the entangled pairs. To further pump the entanglement of the purified pair, a new entangled pair is subsequently created using $m-1$ qubits operated in parallel. The average time it takes to make $j$ rounds of purification is thus estimated as
\begin{equation} \label{eq:pur1}
\tau_{\text{pur},j}=\frac{\tau_{\text{pur},j-1}+\tau_{pur}+\tau_{\text{pair},1:m-1}}{P_{\text{pur}}(F_{j-1},F_{0})},
\end{equation}
with $\tau_{\text{pur},0}=\tau_{\text{pair},2;m}-\tau_{\text{pair},1:m-1}$. Here $F_{j-1}$ is the fidelity of the purified pair after $j-1$ purifications. 
The total rate of a repeater, consisting both of purification and entanglement swapping, depends on the specific repeater achitecture. We will first consider the case of both a parallel and sequential repeater operated with deterministic gates and afterwards the same situations with probabilistic gates.  

\subsection{Deterministic gates}

For a parallel repeater with $n$ swap levels and deterministic gates, we first estimate the average time it takes to generate $2^{n}$ purified pairs, i.e. a purified pair in each elementary link. We assume that each pair is purified $j$ times such that the time to generate one purified pair is
\begin{eqnarray} \label{eq:pur2}
\tau_{\text{pur},j}&=&\frac{\mathcal{Z}_{2;m}(P_{0})(L_{0}/c+\tau_{\text{local}})}{P_{\text{pur}}(F_{0},F_{0})\cdots P_{\text{pur}}(F_{j-1},F_{0})}\nonumber \\
&&+\sum_{i=0}^{j-1}\frac{\tau_{\text{pur}}}{P_{\text{pur}}(F_{i},F_{0})\cdots P_{\text{pur}}(F_{j-1},F_{0})} \nonumber \\
&&+\sum_{i=1}^{j-1}\frac{\mathcal{Z}_{1;m-1}(P_{0})(L_{0}/c+\tau_{\text{local}})}{P_{\text{pur}}(F_{i},F_{0})\cdots P_{\text{pur}}(F_{j-1},F_{0})},
\end{eqnarray}
where we have solved the recurrence in Eq.~\eqref{eq:pur1}. We now wish to estimate the total time, $\tau_{\text{link},2^{n}}$ it takes to make purification in every elementary link, i.e. the time it takes to make $2^{n}$ pairs. A lower limit of $\tau_{\text{link},2^{n}}$ is simply $\tau_{\text{pur},j}$ but this is a very crude estimate if the purification have a limited success probability since the time is not determined by the average time but by the average time for the last link to succeed. We therefore make another estimate of the average time by treating $\tau_{\text{pur},j}$ as consisting of $2j$ independent binomial events with probabilities 
\begin{eqnarray}
P_{1}&=&\frac{P_{\text{pur}}(F_{0},F_{0})\cdots P_{\text{pur}}(F_{j-1},F_{0})}{\mathcal{Z}_{2;m}(P_{0})} \\
P_{2}^{(i)}&=&P_{\text{pur}}(F_{i},F_{0})\cdots P_{\text{pur}}(F_{j-1},F_{0}) \\
P_{3}^{(i)}&=&\frac{P_{\text{pur}}(F_{i},F_{0})\cdots P_{\text{pur}}(F_{j-1},F_{0})}{\mathcal{Z}_{1;m-1}(P_{0})}.
\end{eqnarray}
We then estimate the average time, $\tau_{link,2^{n}}$ it takes to make $2^{n}$ purified pairs as
\begin{eqnarray} \label{eq:purnlink}
\tau_{\text{link},2^{n}}&=&\mathcal{Z}_{2^{n};2^{n}}(P_{1})(L_{0}/c+\tau_{\text{local}}) \nonumber \\
&&+\sum_{i=0}^{j-1}\mathcal{Z}_{2^{n},2^{n}}(P_{2}^{(i)})\tau_{\text{pur}} \nonumber \\
&&+\sum_{i=1}^{j-1}\mathcal{Z}_{2^{n},2^{n}}(P_{3}^{(i)})(L_{0}/c+\tau_{\text{local}}).
\end{eqnarray}
Eq.~\eqref{eq:purnlink} is a better estimate for the average time than $\tau_{\text{pur},j}$ in the limit of small success probabilities since it takes into account that we need success in all links. This is contained in the factors $\mathcal{Z}_{2^{n};2^{n}}$. However, it overestimates the average distribution time when the purification has a large success probability. How much it overestimates depends on $n$ and $j$. Comparing  $\tau_{\text{pur},j}$ to Eq.~\eqref{eq:purnlink} we find numerically that for $n\leq5$ and $j\leq2$ there is a factor $\lesssim2$ between the two estimates, in the limit of large success probability for the purification operation. As discussed in Sec.~\ref{sec:optim} we never consider more than 5 swap levels in our optimization and since we have a limited number of qubits pr. repeater station, we will never have to consider more than 2 rounds of purification. We can therefore use the estimate for $\tau_{\text{link},2^{n}}$ given in Eq.~\eqref{eq:purnlink}.

To get the average time it takes to distribute one entangled pair over the total distance, $L_{tot}$, of the repeater, we need to add the time of the entanglement swapping, $\tau_{\text{swap},nd}$ to $\tau_{\text{link},2^{n}}$. We estimate $\tau_{\text{swap},n}$ as
\begin{equation}
\tau_{\text{swap},n}=(2^{n}-1)L_{0}/c+n\tau_{\text{c}},
\end{equation} 
where the first term is the time of the classical communication and $\tau_{\text{c}}$ is the time of the CNOT operation involved in the swap procedure. The average distribution rate, of a parallel repeater with deterministic gates, is thus $r=1/(\tau_{\text{link},2^{n}}+/\tau_{\text{swap},n})$.

For a repeater using sequential entanglement creation with deterministic gates, we estimate the time it takes to generate purified pairs in all $2^{n}$ pairs as
\begin{equation}
\tau^{(s)}_{\text{link},2^{n}}=\left(\tau_{\text{link},2^{n-1}}\right)\vline_{m\to2m}+\left(\tau_{\text{link},2^{n-1}}\right)\vline_{m\to2m-1}.
\end{equation}
Here we have indicated that the number of qubits, which can be operated in parallel is $2m$ for the first $2^{n-1}$ pairs and $2m-1$ for the next $2^{n-1}$ pair compared to the parallel repeater, where only $m$ qubits can be used in all $2^{n}$ pairs. Note, that we have assumed that first entanglement is established in half of the links and only when this is completed, entanglement is created in the remaining half of the links. This is clearly not the fastest way of operating the repeater but it gives an upper limit of the average distribution time and hence a lower limit on the rate. The entanglement swapping of the sequential repeater is exactly the same as for the parallel repeater and the average total rate, of the sequential repeater with deterministic gates, is thus $r=1/(\tau^{(s)}_{\text{link},2^{n}}+t_{\text{swap},n})$. 

\subsection{Probabilistic gates}

To estimate the total, average distribution time of a repeater using parallel entanglement creation with $n$ swap levels and probabilistic gates, we will again treat $\tau_{\text{pur},j}$ as consisting of $2j$ independent binomial events, as we did for the deterministic gates. The time it takes to make a single swap can be estimated as
\begin{eqnarray} \label{eq:swap1}
\tau_{\text{swap},1}&=&\frac{\mathcal{Z}_{2;2}(P_{1})(L_{0}/c+\tau_{\text{local}})}{P_{\text{swap}}}+\frac{L_{0}/c}{P_{\text{swap}}}+\frac{\tau_{\text{c}}}{P_{\text{swap}}} \nonumber \\
&&+\sum_{i=0}^{j-1}\frac{\mathcal{Z}_{2;2}(P_{2}^{(i)})\tau_{\text{pur}}}{P_{\text{swap}}} \nonumber \\
&&+\sum_{i=1}^{j}\frac{\mathcal{Z}_{2;2}(P_{3}^{(i)})(L_{0}/c+\tau_{\text{local}})}{P_{\text{swap}}},
\end{eqnarray}
where $P_{\text{swap}}$ is the probability of the swap operation, i.e. the probability of the CNOT gate. Eq.~\eqref{eq:swap1} can be iterated such that the average time it takes to make $n$ swap levels is estimated as
\begin{eqnarray} \label{eq:swap2}
\tau_{\text{swap},n}&=&\frac{\tilde{\mathcal{Z}}_{n;1}(P_{\text{swap}},P_{1})(L_{0}/c+\tau_{\text{local}})}{P_{\text{swap}}} \nonumber \\
&&+\sum_{i=1}^{n}\frac{\tilde{\mathcal{Z}}_{n;i}(P_{\text{swap}},P_{\text{swap}})(2^{i-1}L_{0}/c+\tau_{\text{c}})}{P_{\text{swap}}} \nonumber \\
&&+\sum_{i=0}^{j-1}\frac{\tilde{\mathcal{Z}}_{n;1}(P_{\text{swap}},P_{2}^{(i)})\tau_{\text{pur}}}{P_{\text{swap}}} \nonumber \\
&&+\sum_{i=1}^{j-1}\frac{\tilde{\mathcal{Z}}_{n;1}(P_{\text{swap}},P_{3}^{(i)})(L_{0}/c+\tau_{\text{local}})}{P_{\text{swap}}}, \qquad
\end{eqnarray}
where 
\begin{eqnarray}
\tilde{\mathcal{Z}}_{n;i}(P_{\text{swap}},P)&=&\mathcal{Z}_{2;2}\left(\frac{P_{\text{swap}}}{\tilde{\mathcal{Z}}_{n-1;i}(P_{\text{swap}},P)}\right), \\ \tilde{\mathcal{Z}}_{i;i}(P_{\text{swap}},P)&=&\mathcal{Z}_{2;2}(P) \\
\tilde{\mathcal{Z}}_{i;i}(P_{\text{swap}},P_{\text{swap}})&=&1.
\end{eqnarray}
Here $P$ is either $P_{1}, P_{2}^{(i)}$ or $P_{3}^{(i)}$. In the limit of $P_{0},P_{\text{swap}}\ll1$ and assuming no initial purification, Eq.~\eqref{eq:swap2} reduces to the well-known fomula \cite{sangouard3,sangouard2}
\begin{equation}
\tau_{\text{swap},n}=\frac{\left(3/2\right)^{n}\left(L_{0}/c+\tau_{\text{local}}\right)}{P_{0}P_{swap}^{n}},
\end{equation}
since $\mathcal{Z}_{2;2}(P\ll1)\approx3/(2P)$ and the time of local operations in the swaps can be neglected in this limit. However, for higher success probabilities, Eq.~\eqref{eq:swap2} more accurately estimates the average distribution time. The average rate of a parallel repeater with probabilistic gates and $n$ swap levels is then $r=1/\tau_{\text{swap},n}$. From a numerical study, we again find that for $P_{\text{swap}}\approx1$ and $P_{0}\ll1$, Eq.~\eqref{eq:swap2} underestimates the average distribution rate with a factor that increases with the number of swap levels, $n$. However, for $n\leq 5$ and $j\leq2$, we find that this factor is $\lesssim2$. 

The operation of a sequential repeater with probabilistic gates is not straightforward since it is unclear how the sequential generation of entanglement should take place after a failed swap operation. We therefore choose to assume that initially, entanglement is generated in all $2^{n}$ links sequentially. When this is completed the first round of entanglement swapping is performed. If a swap fails, entanglement is restored in a parallel manner in this section, i.e. the sequential operation is only employed in the initial generation of entanglement. Thus if $i$'th swaps fail in the first swap level an extra waiting time of 
\begin{eqnarray} \label{eq:swap01}
&&\mathcal{Z}_{i;i}\left(\frac{P_{\text{swap}}}{\mathcal{Z}_{2;2}(P_{1})}\right)(L_{0}/c+\tau_{\text{local}}) \nonumber \\
&&+\mathcal{Z}_{i;i}(P_{\text{swap}})(L_{0}/c+\tau_{\text{c}}) \nonumber \\
&&+\sum_{k=0}^{j-1}\mathcal{Z}_{i;i}\left(\frac{P_{\text{swap}}}{\mathcal{Z}_{2;2}(P_{2}^{(k)})}\right)\tau_{\text{pur}} \nonumber \\
&&+\sum_{k=1}^{j-1}\mathcal{Z}_{i;i}\left(\frac{P_{\text{swap}}}{\mathcal{Z}_{2;2}(P_{3}^{(k)})}\right)(L_{0}/c+\tau_{\text{local}})
\end{eqnarray}
is needed to restore entanglement in the $2i$'th links in a parallel manner and swap them successfully. Eq.~\eqref{eq:swap01} is very similar to Eq.~\eqref{eq:swap1}, which estimates the time needed for a single swap at the first swap level. Nonetheless, the functions $\mathcal{Z}_{i;i}$, which appears in Eq.~\eqref{eq:swap01} takes into account that we need $i$ successful swaps instead of only a single swap.  
Furthermore, we assume that the swap operations of a swap level is only initiated when all swap operations in the preceeding level have been successful. The average time, it takes for all swap operations in the first level to succeed, is then estimated as
\begin{eqnarray} \label{eq:seq1}
\tau^{(s)}_{\text{swap},1}&=&\sum_{i=0}^{2^{n-1}}P_{\text{swap}}^{2^{n-1}-i}(1-P_{\text{swap}})^{i}\Bigg[ \nonumber \\
&&\mathcal{Z}_{i;i}\left(\frac{P_{\text{swap}}}{\mathcal{Z}_{2;2}(P_{1})}\right)(L_{0}/c+\tau_{\text{local}}) \nonumber \\
&&+\mathcal{Z}_{i;i}(P_{\text{swap}})(L_{0}/c+\tau_{\text{c}}) \nonumber \\
&&+\sum_{k=0}^{j-1}\mathcal{Z}_{i;i}\left(\frac{P_{\text{swap}}}{\mathcal{Z}_{2;2}(P_{2}^{(k)})}\right)\tau_{\text{pur}} \nonumber \\
&&+\sum_{k=1}^{j-1}\mathcal{Z}_{i;i}\left(\frac{P_{\text{swap}}}{\mathcal{Z}_{2;2}(P_{3}^{(k)})}\right)(L_{0}/c+\tau_{\text{local}}) \nonumber \\
&&+(L_{0}/c+\tau_{\text{c}})\delta_{i,0} \Bigg],
\end{eqnarray}
where $\delta_{i,0}$ is the Kronecker delta symbol and $\mathcal{Z}_{0;0}=0$. It is seen that in the limit $P_{\text{swap}}\to1$, Eq.~\eqref{eq:seq1} correctly reduces to $\tau^{(s)}_{\text{swap},1}=L_{0}/c+\tau_{\text{c}}$, which simply is the time of the classical communication of the results of the bell measurements and the time of the local operations. Eq.~\eqref{eq:seq1} can be generalized such that the time it takes to perform the $l$'th swap level is
\begin{eqnarray}
\tau^{(s)}_{\text{swap},l}&=&\sum_{i=0}^{2^{n-l}}P_{\text{swap}}^{2^{n-l}-i}(1-P_{\text{swap}})^{i}\Bigg[ \nonumber \\
&&\mathcal{Z}_{i;i}\left(\!\frac{P_{\text{swap}}}{\tilde{\mathcal{Z}}_{l;1}(P_{\text{swap}},P_{1})\!}\right)(L_{0}/c+\tau_{\text{local}}) \nonumber \\
&&+\sum_{k=1}^{l}\mathcal{Z}_{i;i}\!\left(\frac{P_{\text{swap}}}{\tilde{\mathcal{Z}}_{l;k}(P_{\text{swap}},P_{\text{swap}})}\right)\!(2^{k\!-\!1}L_{0}/c+\tau_{\text{c}}) \nonumber \\
&&+\sum_{k=0}^{j-1}\mathcal{Z}_{i;i}\left(\frac{P_{\text{swap}}}{\tilde{\mathcal{Z}}_{l;1}(P_{\text{swap}},P_{2}^{(k)})}\right)\tau_{\text{pur}} \nonumber \\
&&+\sum_{k=1}^{j-1}\mathcal{Z}_{i;i}\left(\frac{P_{\text{swap}}}{\tilde{\mathcal{Z}}_{l;1}(P_{\text{swap}},P_{3}^{(k)})}\right)(L_{0}/c+\tau_{\text{local}}) \nonumber \\
&&+(2^{l-1}L_{0}/c+\tau_{\text{c}})\delta_{i,0} \Bigg],
\end{eqnarray} 
which can be compared to Eq.~\eqref{eq:swap2} which estimates the time to make a successful swap at the $n$'th level (let $n\to l$ for comparison). Once again the functions $\mathcal{Z}_{i;i}$ takes into account that we need $i$'th successful swaps instead of just a single successful swap. The total rate of a sequential repeater with probabilistic gates and $n$ swap levels can then be estimated as $r=1/(\tau^{(s)}_{\text{link},2^{n}}+\tau^{(s)}_{\text{swap},1}+\cdots+\tau^{(s)}_{\text{swap},n})$.

\end{document}